\documentclass[times,11pt]{article}
\usepackage{setspace}
\usepackage[left=2.5cm,top=3cm,right=2.5cm,bottom=3cm,bindingoffset=0.5cm]{geometry}

\usepackage{moreverb}

\usepackage{amsmath, amsfonts, epsfig, xspace, bm, mathrsfs}

\usepackage{bm}
\usepackage{subcaption}
\usepackage{mathrsfs}

\usepackage{authblk}

\newcommand\BibTeX{{\rmfamily B\kern-.05em \textsc{i\kern-.025em b}\kern-.08em
T\kern-.1667em\lower.7ex\hbox{E}\kern-.125emX}}

\def\volumeyear{2016}

\newcommand{\bu}{\bm{u}}
\newcommand{\bv}{\bm{v}}
\newcommand{\bnu}{\bm{\nu}}

\newcommand{\bx}{\bm{x}}
\newcommand{\bg}{\bm{g}}

\title{Stable and accurate interface capturing advection schemes
}

\author[1]{Florian De Vuyst\thanks{E-mail: devuyst@cmla.ens-cachan.fr}}
\author[1]{Marie B\'echereau}
\author[1,2,3]{Thibault Gasc}
\author[3]{Renaud Motte}
\author[4]{Mathieu Peybernes}
\author[5]{Raphael Poncet}

\affil[1]{CMLA, ENS Cachan, CNRS, Universit\'e Paris-Saclay, 94235 Cachan France}
\affil[2]{Maison de la simulation, USR 3441, CEA Saclay, 91191 Gif-sur-Yvette, France}
\affil[3]{CEA DAM DIF, F-91297 Arpajon France}
\affil[4]{CEA Saclay, DEN, 91191 Gif-sur-Yvette, France}
\affil[5]{CGG, 27 avenue Carnot, 91300 Massy France}

\begin{document}
\maketitle

\begin{abstract}
In this paper, stable and ``low-diffusive'' multidimensional interface capturing (IC) schemes using slope limiters are discussed. It is known that direction-by-direction slope-limited MUSCL schemes
create geometrical artefacts and thus return a poor accuracy. We here focus on this particular issue and show that the reconstruction of
gradient directions are an important factor of accuracy. The use of a multidimensional limiting process (MLP) added with an adequate time integration scheme leads to an artefact-free and instability-free
interface capturing (IC) approach. Numerical experiments like the reference Kothe-Rider forward-backward advection case show the accuracy of the approach. We also show that the approach can be extended to
the more complex compressible multimaterial hydrodynamics case, with potentially 
an arbitrary number of fluids.
We also believe that this approach is appropriate for multicore/manycore architecture because of
its SIMD feature, which may be another asset compared to interface reconstruction approaches.
\end{abstract}


\maketitle

\begin{flushleft}

\end{flushleft}
\vspace{-6pt}

\section{Introduction and motivation}
%
\medskip
Nowadays there are recognized computational methods to numerically simulate material interfaces or moving free boundaries. Among the well-known approaches, let us
mention the levelset approach pioneered by Osher-Sethian \cite{Osher}, the family of interface reconstructions (IR) algorithms \cite{Caboussat, Breil} that can be more or less sophisticated
(moments-of-fluid approaches~\cite{Breil} being the most sophisticated),
and diffuse interface capturing (IC) methods \cite{Thuburn,Ubbink99,Despres2001,Despres2007,Despres2010,Bokanowski,Michel,Champmartin,Khrabry,Faucher,Muzaferija,Qian}. Each of these methods both show advantages and drawback, and
thus are more or less suitable for different problems according to the kind of application, the expected properties and the quantities
of interest to compute. For example, if conservation properties are mandatory,
levelset methods are not the right candidate family: even if today we find
volume conservative levelset methods, mass conservation may not be strictly fulfilled, 
what can be not accurate enough for some highly compressible flows or
flows with high ratios of density. 
After decades of sustained developments and research in this fields, interface capturing (IC) methods are still an active field of investigation 
(see the recent references~\cite{Despres2001,Despres2007,Despres2010,Bokanowski,Champmartin,Park2011,Qian,Yoon2009}). They can show
advantages like a natural extension to an arbitrary number of fluids, phases of materials, the possibility to deal with complex topologies and configurations (triple points, ...) in a rather easy way, the simplicity of code development,
debugging and optimization. \medskip

There is also another feature not to forget: the compute performance of the methods, especially
for multicore or manycore parallel computer architectures that appear to be the current
and future driving hardware trend. Multicore/manycore processors allow of 
speedups only if algorithms are suitable for that. Key factors of 
performance are typically a well-balanced operational intensity, data
coalescence in memory, cache blocking, processor occupancy and 
SIMD\footnote {SIMD = single instruction, multiple data} feature of the 
algorithms \cite{Poncet}.
We advocate the idea that today a good computational approach is a good trade-off between numerical analysis requirements (like order of accuracy and stability), generalization/flexibility property and fast or practical implementation features. 
Our experience in computational methods, performance evaluation and performance modeling \cite{Poncet} make
us believe that interface capturing schemes are good candidates for these performance issues, ready to fulfil most of these performance factors.\medskip

As a summary, we look for interface capturing methods that share the following properties:
\begin{itemize}
\item numerical stability and low-diffusive aspect;
\item SIMD-type algorithms;
\item simple coding and debugging;
\item relatively easy extension to an arbitrary number of materials;
\item relatively  easy extension to three-dimensional problems;
\item natural extension to unstructured meshes;
\item relatively simple prediction of computing performance.
\end{itemize}
This paper more focuses on stability, low-diffusive feature and accuracy 
of interface capturing methods, but we keep in mind all the above requirements.
The paper is organized as follows. In section~2, we will test standard MUSCL
slope limiter-based finite volume schemes for interface capturing and will show a set
of artefacts and instabilities by numerical evidence. In section 3, we will provide
some elements of analysis about these artefacts. We will enumerate the
requirements for a good interface capturing scheme in section~4 and then
discuss multidimensional limiting process methods in section~5. Numerical
experiments in section 6 will confirm both stability and accuracy of our
approach and considerations for the extension toward compressible multimaterial
hydrodynamics will be given in section~7.   
%
%
\paragraph{Mathematical setup and notations.} We need to describe free boundaries
moving into a continuous medium described in a bounded spatial domain $\Omega\subset\mathbb{R}^d$, $d\in\{2,3\}$. 
Let us denote by~$\bu$ the vector field on the underlying transport phenomenon.
As a first step and for simplicity let us assume that the vector field
only depends on space~$\bx$ and not on time~$t$. We add the following
regularity assumptions on the vector field~:
$\bu\in [W^{1,\infty}(\Omega)]^d$ and~$\bu$ is divergence-free, i.e.
$\nabla \cdot\bu=0$ almost everywhere. Then any quantity $z$ solution the 
pure advection equation
\[
D_t z = \partial_t z + \bu\cdot\nabla z = 0,
\]
is also solution of the conservation law
\[
\partial_t z + \nabla\cdot (z \bu) = 0.
\]
Variable $z$ is a conservative variable in the sense that
\begin{equation}
\frac{d}{dt}\int_{V_t} z(\bx,t)\, d\bx = 0
\end{equation}
for any measurable set $V_t$ which is transported itself by the vector field $\bu$
so that the ``mass'' quantity 
\[
m_t = \int_{V_t} z(\bx,t)\, d\bx 
\] 
is conserved through time $t$.
Because we are dealing with the capture of moving free boundaries, we here consider
discontinuous solutions~$z$ with values only in $\{0,1\}$ at the continuous level.
Let us remark that, from the Reynolds transport theorem,
\[
\frac{d}{dt}|V_t| = \frac{d}{dt}\int_{V_t}1\, d\bx = 
\int_{V_t} \{\partial_t 1 + \nabla\cdot\bu\}\, dx = 0,
\]  
thus there is volume conservation for a divergence-free velocity field. 
So conservation of volume and conservation of ``mass'' are equivalent in this case.
At the discrete level, we want to keep the conservation properties, so 
finite volumes methods are the natural candidates for that. To set the ideas,
let us consider a two-dimensional rectangular bounded domain $\Omega$ and a cartesian discretization mesh with uniform mesh steps $h_x=h_y=h$ leading to square finite volumes.
Let us use $K$ as generic notation of a given finite volume, $A$ an edge of volume $K$,
the vector~$\bnu_A$ being the outer normal unit vector to $A$ pointing out of $K$. We look for finite volume schemes (here written in semi-discrete form -- time discretization will be discussed later) in the form
\begin{equation}
\frac{d z_K}{dt} = - \frac{1 }{|K|}\sum_{A\subset \partial K} |A|\, z_A\, \bu_A\cdot\bnu_A
\end{equation} 
where $z_A$ (resp. $\bu_A$) denotes a certain value of $z$ (resp. $\bu$)
at the midpoint of edge $A$. The quantity
\[
\text{div}^h_K (\bu z) = \frac{1 }{|K|}\sum_{A\subset \partial K} |A|\, z_A\, \bu_A\cdot\bnu_A
\]
is nothing else but a discrete divergence operator of the vector flux $\bm{f}=\bu z$
into the cell $K$. 
Because we deal with divergence-free velocity vectors, we may chose $(\bu_A\cdot\bnu_A)$
in order to get the discrete equivalent
\[
\sum_{A\subset \partial K} |A| \bu_A\cdot\bnu_A = 0.
\] 
For that it is sufficient to consider mean values $\bu_A\cdot\bnu_A$ computed as
\[
\bu_A\cdot\nu_A = \frac{1}{|A|} \int_A \bu\cdot\bnu_A\, d\sigma.
\]
In summary, we have defined a numerical normal flux
\[
\phi_A = (\bu_A \cdot \bnu_A)\, z_A 
\]
across the edge $A$ in the direction $\bnu_A$. We now have only to correctly chose and compute the edge quantity~$z_A$ in order to get a stable and low-diffusive time advance scheme.
\section{Testing MUSCL slope limiter-based schemes for interface transport}
%
\subsection{One-dimensional case}
%
Here we begin with a short introduction and summary of well-known second-order finite volume solvers
using MUSCL reconstruction for the one-dimensional advection equation
\[
\partial_t z + \partial_x (u z) = 0
\]
for a given constant real number $\bu\neq 0$. Spatially semi-discrete conservative schemes
for a uniform grid read
\begin{equation}
\frac{d z_j}{dt} = -  \frac{\psi_{j+1/2}-\psi_{j-1/2}}{h}, \quad\quad j\in\mathbb{Z},
\end{equation}
where $h$ denotes a constant spatial step of the grid $\{x_j=jh\}_j$, $j\in\mathbb{Z}$,
and the quantities~$\psi_{j+1/2}$ are numerical fluxes between cells $j$ and $(j+1)$. 
The fluxes are at least consistent with
the physical (linear) flux $f(z)=uz$ to get first-order accuracy. Slope-based MUSCL
reconstructions methods try to locally reconstruct a slope for each cell by finite differences
and then limit the slopes for total variation diminishing (TVD) stability 
purposes \cite{Godlewski,Toro}. 
Without time discretization, the upwind numerical fluxes generally are written in viscous form
\begin{equation}
\psi_{j+1/2} = \frac{f(z_{j+1/2}^-)+f(z_{j+1/2}^+)}{2} - \frac{1}{2} |u| 
(z_{j+1/2}^+-z_{j+1/2}^-),
\label{eq:10}
\end{equation}
where $z_{j+1/2}^-$ and $z_{j+1/2}^+$ are left and right extrapolation values at location
$x_{j+1/2}=(j+1/2)h$ respectively, according to a conservative piecewise linear approximation
\[
z(x) = z_j + s_j (x-x_j), \quad\quad x\in (x_{j-1/2},x_{j+1/2}),
\]
and then
\[
z_{j+1/2}^- = z_j + \frac{h}{2} s_j,\quad 
z_{j-1/2}^+ = z_j - \frac{h}{2} s_j.
\]
The numerical flux~\eqref{eq:10} can be rewritten
\[
\psi_{j+1/2} = \left\{\begin{array}{l}
u\, z_{j+1/2}^- \quad \text{if } u\geq 0,\\ [1.1ex]
u\, z_{z+1/2}^+ \quad \text{if } u\leq 0
\end{array}\right.
\]
and clearly shows the underlying upwind process.
In the slope limitation theory, one tries to reconstruct and limit a slope according to
some requirement like second-order accuracy and TVD property for the scheme. Cell-centered 
three-point slope reconstructions have the form
\[
s_j = \frac{1}{h}\, \phi(\Delta z_j^-, \Delta z_j^+) 
\]
with $\Delta z_j^-=z_j-z_{j-1}$, $\Delta z_j^+=z_{j+1}-z_j$, and $\phi$ is a limiter
function. Among, standard limiter functions $\phi$, let us mention the ``compressive'' ones that
will be used in the sequel of the paper:
\begin{itemize}
\item Superbee limiter~\cite{Roe}:
\[
\phi^S(a,b) = \text{sgn}(a)\, (ab\geq 0)\, \max[\min(|a|,2|b|), \min(2|a|,|b|)];
\]
\item a more compressive limiter (referred to as the ``overbee'' limiter in this paper,
used for example in~\cite{Qian}):
\[
\phi^O(a,b) = 2\,\text{sgn}(a)\, (ab\geq 0)\, \min(|a|,|b|);
\]
\end{itemize}
Superbee limiters allow for second-order reconstruction because of the property $\phi^S(a,a)=a$ and regularity whereas the ``overbee'' limiter overestimates the slopes:
\[
\phi^O(a,a) = 2a\quad \forall a.
\]
The overbee limiter is a priori intended for use with step-shaped or staircase functions only. Although being overcompressive, it leads to a $L^\infty$-stable scheme under
CFL conditions less than~$1/2$. \medskip

For full discretized numerical schemes (by the explicit Euler scheme for example), there
are also CFL-dependent limiter functions, like Roe's Ultrabee limiter \cite{Roe}, depending on
the Courant number $\nu=|u|\Delta t/h$, with $\Delta t$ as time step. Despr\'es
and Lagouti\`ere \cite{Despres2001} in their construction of the most compressive stable scheme for interface
advection (known as limited-downwind approach) have reinterpreted their construction in terms of flux
limiter and then retrieve the Ultrabee limiting process. \medskip

%
%
As an illustration, let us consider the simple test case of transport of the
initial top hat function 
\[
z^0(x) = 1_{(0\, \leq\,  x\, \leq\,  1/2)}(x)
\]
on the interval $\Omega=(0,1)$ with periodic boundary conditions, $u=1$. 
Let us use a uniform grid made of 250 points. We implement the MUSCL scheme using the overbee limiter with the explicit Euler time discretization. We use a Courant number $\nu=0.35$. Final time is~$t=5/4$. On figure~\label{fig:0}, the discrete solution
at final time is plotted. We can observe a quite good capture of the discontinuities
``at the eye norm''. Looking more deeply at the viscous profile by visualizing
the logarithm of the quantity $z(1-z)$ that is representative of the smearing rate,
one can observe a smearing of about 10 points that decreases in log scale with a cutoff at~$10^{-10}$. Than can be be simply explained with the following
configuration: consider $u>0$, discrete values~$z_{j-1}^n=1$, $z_j^n<1$ close to 1 and $z_{j+1}^n$ 
close to 0. Applying the MUSCL/overbee limiter under this configuration
clearly gives 
\begin{eqnarray*}
z_j^{n+1}-1 &=& z_j^n-1 - \nu ([z_j^n + \frac{1}{2}\, 2(z_j^n - z_{j-1}^n)] - 1) \\[1.1ex]
            &=& (1-2\nu) (z_j^n-1).
\end{eqnarray*}
Stability is ensured under the CFL condition $\nu\leq 1/2$, but for $\nu<1/2$, the have
a geometric series with factor $(1-2\nu)\in (0,1)$. 
The convergence rate toward 1 is CFL-dependent and can be rather small for small Courant values. The only way to avoid this is to use compressive CFL-dependent limiters like the Ultrabee one and is at the source of Roe's construction \cite{Roe} or Despr\'es-Lagouti\`ere in their construction of limited-downwind antidiffusive approach \cite{Despres2001}.
Another important thing to notice in this experiment is the asymptotic bound in time of the smearing rate, i.e.
\[
t\mapsto \max_{s\in(0,t)}\, \int_0^1 [z(1-z)](s)\, ds
\]
is bounded (figure~\ref{fig:0}b), which states in some sense the low-diffusive
feature of the approach.
\begin{figure}[ht]
\begin{center}
\begin{subfigure}{0.4\textwidth}
\centering\includegraphics[height=0.2\textheight]{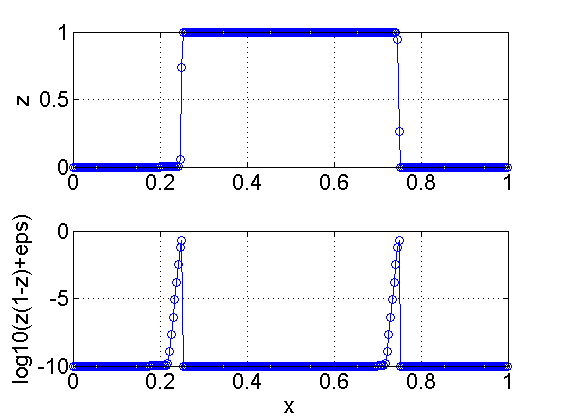}
\caption{Profile of $z$ and 
$\log_{10}(z(1-z)+10^{-10})$ at final time.}
\end{subfigure}
	\begin{subfigure}{0.4\textwidth}
	\begin{center}
	\includegraphics[height=0.2\textheight]{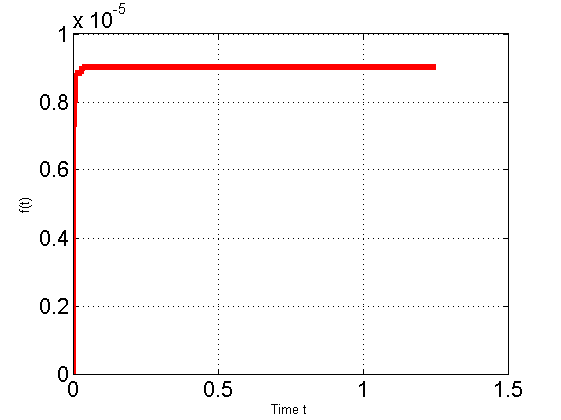}
	\caption{History of quantity \\
	$t\mapsto \max_{s\in(0,t)}\, \|[z(1-z)](s)\|_{L^1(0,1)}$.}
	\end{center}
	\end{subfigure}
\end{center}
\caption{Numerical experiment: assessment of numerical diffusion and smearing
of step-shaped functions when the overbee limiter is used. One can observe a smearing
of a few points ``at the eye norm'', and about $O(10)$ point in log scale. Time evolution
of the smearing rate $t\mapsto \max_{s\in(0,t)}\, \int_0^1 (z(1-z))(x,s)\,dx$ shows an asymptotic bound.} \label{fig:0}
\end{figure}

\subsection{Artefacts and instabilities encountered for multidimensional problems} \label{sec:22}
%
For multidimensional problems, the state of theoretical analysis of slope limiters is still
nowadays quasi-open or quite poor. Total variation theory for example cannot be  extended for multidimensional
problems. Anyway, we would like to experimentally observe the behavior of direction-by-direction
slope limiters on a natural multidimensional extension of the MUSCL reconstruction upwind scheme.

\begin{figure}[ht]
\begin{center}
\begin{subfigure}{0.45\textwidth}
\includegraphics[width=\linewidth]{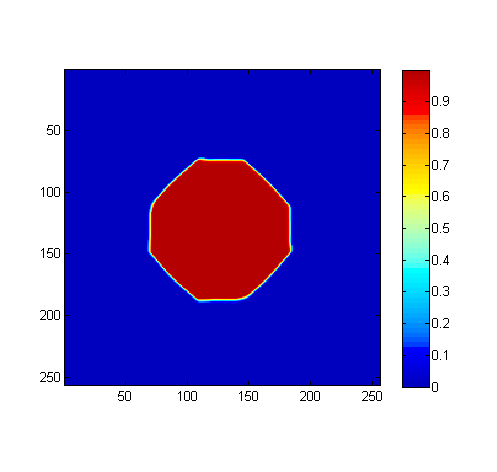}
\caption{Colored representation of the solution $z$.}
\end{subfigure}
\begin{subfigure}{0.45\textwidth}
\includegraphics[width=\linewidth]{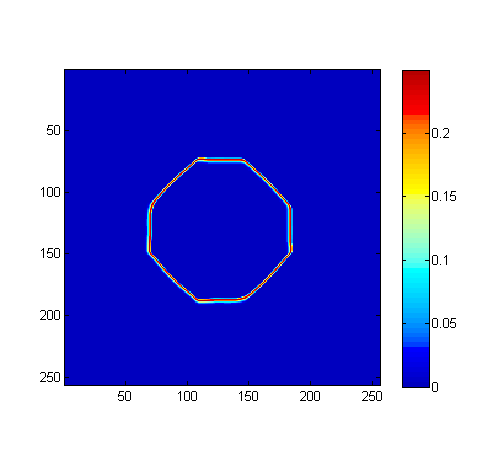}
\caption{Colored representation of  $z(1-z)$.}
\end{subfigure}
\end{center}
\caption{Disk advected into a uniform vector field $\bu=(1,1)$ in the domain
$[-1,1]^2$ (periodic boundary conditions) which evolves toward an non-physical octagon-shaped form. Clearly
the error is $O(1)$. Discrete solution at final time $t=10$, cartesian grid $256^2$. Slope limiters are direction-by-direction superbee limiters.
RK2 is used as time integrator and the CFL number is $0.3$.} \label{fig:5}
\end{figure}

Let us define three ``simple'' advection cases.  First, consider the square domain
$\Omega=(-1,1)^2$ discretized with a cartesian grid made square of edge length $h$.
Consider also periodic boundary conditions. As initial data, let us define a disk-shaped
hat function
\[
z(\bx,t=0) = 1_{(x^2+y^2<0.2)}(\bx).
\]
and consider a uniform velocity field generated by the diagonal advection vector $\bu=(1,1)$. 
For the advection scheme, consider the upwind MUSCL approach with the direction-by-direction
slope limiter Superbee and a Runge-Kutta RK2 time integration for the time advance scheme.
We use a CFL number of 0.2\,.
Final time of numerical simulation is $t=10$, the disk-shaped function should be retrieved
at final time at its initial location. The results are plotted in figure~\ref{fig:5}.
We observe that the disk interface degenerates and artificially evolves in time toward
an octagon-shaped boundary, as already noticed by Despr\'es and Lagouti\`ere 
in~\cite{Despres2007} when
the limited-downwind anti-dissipative approach is used. The numerical scheme does not create new extrema and is locally
monotonicity-preserving, but some accuracy is lost and it is disappointing to state that
the error is $O(1)$ at final time. Some error is accumulating in time.
Let us also claim that a directional splitting strategy does not solve this artefact problem.
\medskip

\begin{figure}[ht]
\begin{center}
\begin{subfigure}{0.45\textwidth}
\includegraphics[width=\linewidth]{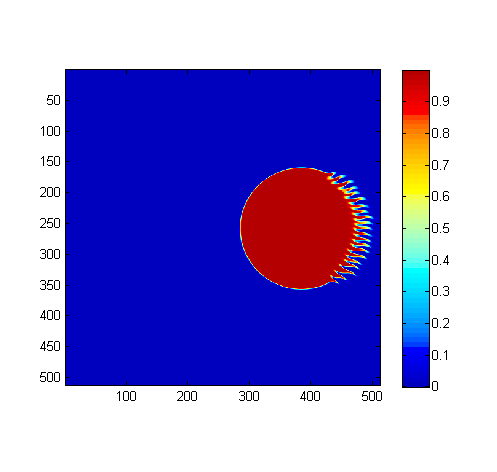}
\caption{Colored representation of the solution $z$.}
\end{subfigure}
\begin{subfigure}{0.45\textwidth}
\includegraphics[width=\linewidth]{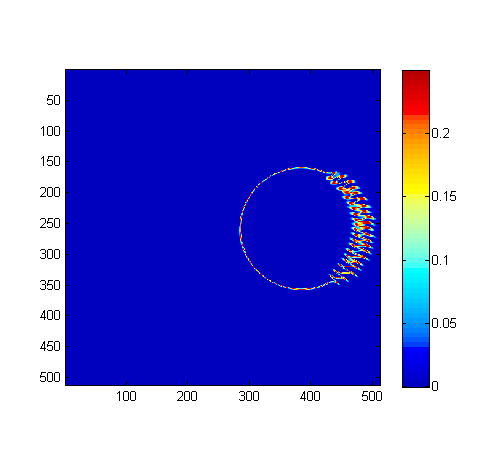}
\caption{Colored representation of  $z(1-z)$.}
\end{subfigure}
\end{center}
\caption{Pure rotation of an disk into a rotating velocity vector field
$\bu(\bx)=(-y,x)$.  Discrete solution at final time $t=2\pi$, cartesian grid $512^2$.
One can observe spurious ``zigzag'' instabilities at the interface.
Slope limiters are direction-by-direction Superbee limiters.
The Euler explicit scheme is used as time integrator and the CFL number is $0.3$.}
\label{fig:rotation}
\end{figure}

As a second test case, let us still consider the same geometry with periodic boundary conditions,
but now with a pure rotating velocity field defined by $\bu(\bx)=(-y,x)$. As initial
condition, consider the disk-shaped function
\[
z(\bx,t=0) = 1_{\left((x-\frac{1}{2})^2+y^2 \, <\, 0.15\right)}(\bx).
\]
Now the computational grid is $512^2$. We still use the MUSCL approach with the Superbee limiter and an explicit Euler time integration. We use a CFL number of 0.3.
Results are shown in figure~\ref{fig:rotation} at final time $t=2\pi$ corresponding to
a complete revolution of the disk in the domain. One can observe some 
free boundary zigzag-shaped instabilities in region where the local Courant
number is rather large. Zigzag instabilities are a recurrent problem already
reported in the literature for interface capturing methods \cite{Michel}. A way to fix
the problem is to use far smaller CFL numbers, but of course at the price of
a weaker performance and a greater diffusivity. \medskip

\begin{figure}[ht]
\begin{center}
\begin{subfigure}{0.45\textwidth}
\includegraphics[width=\linewidth]{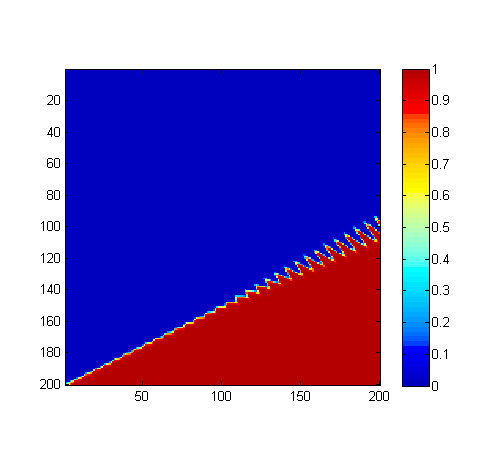}
\caption{Colored representation of the solution $z$.}
\end{subfigure}
\begin{subfigure}{0.45\textwidth}
\includegraphics[width=\linewidth]{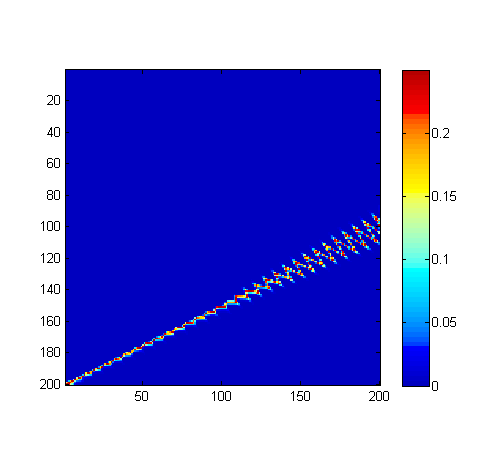}
\caption{Colored representation of  $z(1-z)$.}
\end{subfigure}
\end{center}
\caption{Steady discontinuity problem. Expected steady discontinuous solution
with interface aligned with the velocity direction $(\bu\cdot\nabla z=0)$.
For rather big values of Courant numbers (less than one), one can observe the emergence of zigzag
instabilities, even when the expected discrete steady state is used as initial data. 
Limiter function here is the so-called compressive overbee limiter. We believe that the source of instability is due to a bad evaluation of the gradient direction, creating
over/under-evaluations of the directional convective fluxes.}\label{fig:steady}
\end{figure}

Finally, consider a third case of steady discontinuity aligned with
the uniform velocity vector field (but not with the grid directions). Consider the spatial domain $\Omega=(0,1)^2$ discretized by a Cartesian grid
$200\times 200$ with Neumann boundary conditions, a uniform velocity field generated by the vector $\bu=(2,1)$ and an initial data made up of a discontinuity aligned with the velocity field~:
\[
z(\bx,t=0) = 1_{(y\, \leq \, x/2)}(x,y).
\]
The initial data is projected over piecewise constant functions, with 
``mixed cells'' that discretize the interface. 
Because of the velocity vector $\bu=(2,1)$, $z$ values into mixed cells actually 
are $1/4$ or $3/4$ (see also the next section for more details). 
The CFL number is 0.25 and the overbee limiter is used. As time discretization,
of the first-order explicit Euler scheme is used. Final time is $t=2$. As observed on figure~\ref{fig:steady}, the initial discontinuity becomes unstable and zigzag
modes appear again. Zigzag instabilities tend to be amplified downstream and moreover
produce a spurious unstationary field.

\section{Elements of analysis}
%
\subsection{Effects of one-dimensional slope limiters}
%
From the results of the previous section, as a summary we observe two kinds
of spurious solutions: i) stable evolution toward a non-physical solution (no instability)
with privileged directions (grid-aligned directions, diagonal);
ii) appearance of zigzag-shaped interface instabilities. 

Regarding i), there is a loss of accuracy in this case. Of course the analysis of the
determination of the  order of accuracy is quite hard to achieve because the loss of accuracy of course occurs at locations where the solution is exactly non-smooth (actually discontinuous). 
In any event, let us try to find some expected consistency or accuracy conditions in particular cases.
%
%
\begin{figure}[ht]
\begin{center}
\includegraphics[height=0.15\textheight]{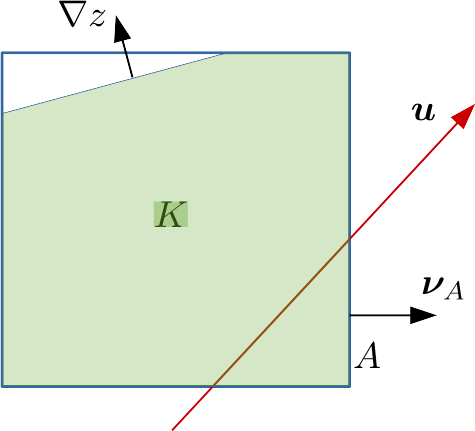}
\caption{Geometrical elements and privileged directions for the interface advection problem} \label{fig:geo}
\end{center}
\end{figure}
In the discrete advection problem, we have privileged directions 
(see figure~\ref{fig:geo}): first
the unit vector~$\bnu_A$ which is normal to the edge~$A$, then the direction 
of advection (given by~$\bu$) and the gradient~$\nabla z$. \medskip

Let us consider a nontrivial uniform velocity field 
$\bu(\bx)=\text{Cst}=\bu$, $|\bu|\neq 0$, and a stationary solution $z$
of the problem. Then $z$ is solution of $\bu\cdot\nabla z = 0$, or
\[
\nabla\cdot(\bu z) = 0 
\]
in conservative form.
If a stationary interface exists, then it is parallel to the line of velocity directions.
Let us see what happen at the discrete level. Recall the discrete
divergence operator
\[
\text{div}^h_K (\bu z) = \frac{1 }{|K|}\sum_{A\subset \partial K} |A|\, z_A\, \bu_A\cdot\bnu_A.
\]
For a uniform velocity field $\bu$, this reduces to
\begin{equation}
\text{div}^h_K (\bu z) = \bu\cdot \left( \frac{1 }{|K|}\sum_{A\subset \partial K} |A|\, z_A\, \bnu_A\right).
\label{eq:2}
\end{equation}
In the above expression we recognize a discrete gradient operator for $z$~:
\[
\text{grad}_K^h(z) = \frac{1 }{|K|}\sum_{A\subset \partial K} |A|\, z_A\, \bnu_A.
\]
%
Assume that locally we have the linear reconstruction
\begin{equation}
\mathscr{I} z(\bx) = z_K + \bg_K\cdot (\bx-\bx_K). 
\label{eq:3}
\end{equation}
Putting~\eqref{eq:3} into~\eqref{eq:2} gives us
\[
\text{div}^h_K (\bu z) = \bu\cdot \left( \frac{1 }{|K|}\sum_{A\subset \partial K} |A|\, 
(z_K+\bg_K\cdot(\bx_A-\bx_K))\, \bnu_A\right).
\]
From the geometrical property
\[
\sum_{A\subset\partial_K} |A|\bnu_A = \bm{0},
\]
we get
\[
\text{div}^h_K (\bu z) = \bu\cdot\left(\frac{1 }{|K|}\sum_{A\subset \partial K} |A|\, 
\bnu_A (\bx_A-\bx_K)^T\right) \bg_K.
\]
By applying Green's formula, it is easy to check that
\[
\frac{1 }{|K|}\cdot\sum_{A\subset \partial K} |A|\, 
\bnu_A (\bx_A-\bx_K)^T = \bm{I},
\]
so that we find
\[
\text{div}^h_K (\bu z) = \bu\cdot\bg_K.
\]
The question is to know whether $\bu\cdot\bg_K$ is close to zero or not. Of course,
it depends on how~$\bg_K$ is computed, and in particular it depends on the choice of slope limiters. We have the following consistency results:
\begin{enumerate}
	\item if $\bg_K \propto \nabla z$, then we have $\text{div}^h_K (\bu z)=0$
	exactly, without error;
	\item if $|\bg_K\cdot \bu|=O(h)$ then we have a first-order consistent formula;
	\item if for some reason, there exists a constant $\beta>0$ which is independent
	of $h$, such that $|\bg_K\cdot \bu|>\beta$, we have a $0$-order consistency. 
\end{enumerate}
We claim that case 3) is exactly what occurs when direction-by-direction 
one-dimensional compressive slope limiters are used. In fact, one-dimensional
slope limiters are used to limit the amplitude of gradients not to create 
new extrema and reproduce a derivative for smooth solutions, but may produce
incorrect gradient directions in the multidimensional case and for non-smooth solutions.

\begin{figure}[ht]
\begin{center}
(a)\includegraphics[width=0.3\textwidth]{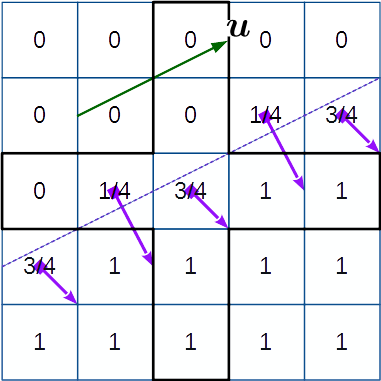}\quad\quad
(b)\includegraphics[width=0.3\textwidth]{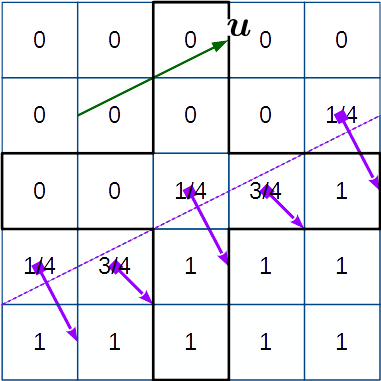}
\caption{Computation of $\text{div}_K^h(\bu z)$ for the following discrete
solutions, using direction-by-direction slope limiters (superbee or ultrabee)
at the center of each grid.
The 9-point cross-shaped stencil for gradient reconstruction is plotted. Values
of $z$ are given for each finite volume. The continuous interface is drawn in dashed
line and is parallel to the uniform velocity vector $\bu=(2,1)$. 
In case b) the reconstructed upwind MUSCL fluxes return the expected value $\text{div}_K^h(\bu z)=\bu\cdot(1/2,-1)/h=0$ while in case a) we find an incorrect value $\text{div}_K^h(\bu z)=\bu\cdot(1/2,-1/2)/h=\frac{1}{2h}=O(h^{-1})$. Let us also remark 
the spurious computation of zigzag-like gradient directions along the interface, which is be probably another source of numerical instability.}
\label{fig:2}
\end{center}
\end{figure}

In fact, reality is a little bit more tricky because reconstructed MUSCL upwind schemes
also take into account the direction of advection by some upwind process.
Let us consider the example given into figure~\ref{fig:2}. Let us still
consider a uniform velocity field $\bu=(2,1)^T$ and a continuous stationary interface
aligned with $\bu$. 
To derive a discrete solution, we project the continuous solution
on piecewise constant solutions as suggested by the finite volume theory. One finds
``mixed'' cells with values either $1/4$ or $3/4$ that represent the interface
at the discrete level. Let us assume the use of direction-by-direction slope limiter of type
superbee (or ultrabee that would return the same result). 
The evaluation of the discrete divergence $\text{div}_K^h(\bu z)$ involves a 9-point
cross-shaped stencil as drawn in figure~\ref{fig:2}.
On figure \ref{fig:2}a), we focus on a mixed cell with value $3/4$.
It is not hard to check that the discrete gradient at the center cell (without
upwinding) is in this case $\bg_K=\frac{1}{h}(1/2,-1/2)$ which is already a bad 
gradient. Taking into account the upwinding process, we find that
\[
\text{div}_K^h(\bu z)=\bu\cdot(1/2,-1/2)^T/h=\frac{1}{2h}
\]
This is clearly a bad value (it is expected to find 0), but moreover the value
behaves like~$O(h^{-1})$~! At the neighbour mixed cell with discrete value $1/4$
(figure~\ref{fig:2}b)), the reconstructed gradient is 
$\bg_K=\frac{1}{h}(1/2,-1/2)$, still incorrect. But, curiously, because of the upwind
process, one can check that the resulting discrete divergence has the good value:
\[
\text{div}_K^h(\bu z)=\bu\cdot(1/2,-1)^T/h=0.
\]
Clearly, this example shows an imbalance of grid-aligned directional
derivatives due to nonlinear limitations of the slope limiters, and the
process is unable to predict the gradient directions correctly.

Still on this example, let us also mention the observed oscillatory 
behaviour of discrete gradients along with the interface line, which,
is, in our opinion, probably another source of odd-even interface instability.
%
%
\subsection{Behaviour of compressive limiters for interface normal vectors orthogonal to the velocity}
%
Moreover, we claim that compressive limiters produce a loss of interface geometry 
accuracy for regions where $\nabla z\cdot\bu$ 
is close to zero, i.e. when $\nabla z$ is almost orthogonal to $\bu$, but not exactly.

\begin{figure}[ht]
\begin{center}
\includegraphics[height=0.25\textheight]{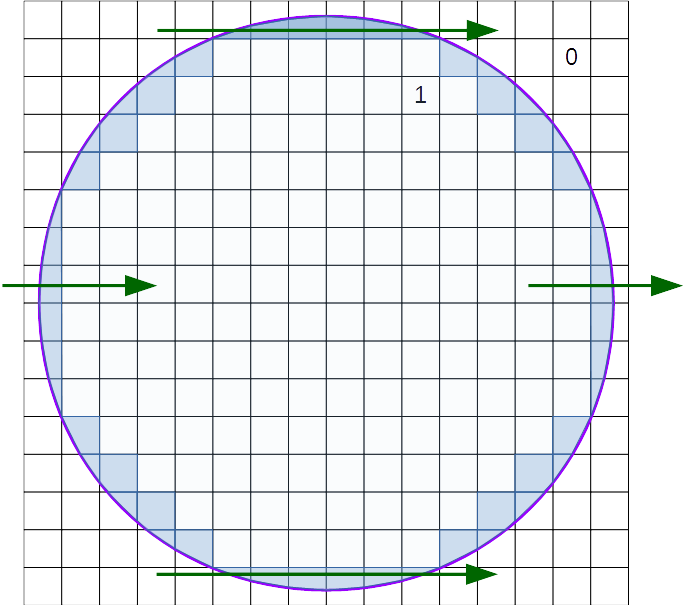}
\end{center}
\caption{Mixed cells (in dark blue) when a disk-shaped solution is projected over piecewise
constant solutions. For locations where $\nabla z$ is almost collinear to the velocity vector (in green), the discrete interface is essentially spread over one point whereas for areas with $\nabla z$ orthogonal or quasi-orthogonal to $\bu$, the discrete $z$ profile in the velocity direction can be rather seen as a smooth varying function with values in $[0,1]$ and thus should be treated as a smooth function in this direction.}
\label{fig:3}
\end{figure}

As represented in figure~\ref{fig:3} with a disk interface as example, the discrete
profile of~$z$ function in regions where $\nabla z\cdot\bu\approx 0$ (top and bottom
of the disk) and in the direction of velocity can be seen as a discretization of a smooth function, varying into $[0,1]$ because of the finite volume mean process. 
On figure~\ref{fig:3} where $\bu$ is aligned with the $x$-axis, applying one-dimensional
too compressive limiters (superbee, overbee, ultrabee) will create spurious staircase-shaped functions in the top and bottom disk areas. Consequently, some of the initial geometry
information will be lost. We believe that that this loss of accuracy is at the origin
of spurious strange attractor shapes like octagons.
To remedy this problem, one could imagine an hybrid slope limiter strategy that takes into account the direction deviation between $\nabla z$ and velocity $\bu$. This
is already proposed for example by Zhang et al.~\cite{Zhang2014} in their
so-called m-CICSAM method.
\subsection{Influence of time discretization} 
%
Time discretization is also an important topic for interface capturing, not only
for accuracy purposes but also especially for stability reasons. 
Time advance schemes for interface capturing are explicit. The simplest one is
the one-step first-order Euler scheme that reads for the semi-discrete (in time) 
advection equation
\[
\frac{z^{n+1} - z^n}{\Delta t} + \nabla \cdot (\bu z^n) = 0. 
\] 
It is very easy to derive the equivalent equation of the explicit first-order Euler scheme, i.e. the equation which is solved at the second-order accuracy by the Euler scheme,
by Taylor expansions. One finds
\[
\partial_t z + \nabla \cdot (\bu z) 
= \frac{\Delta t}{2}\, \nabla\cdot(-(\bu\otimes\bu)\, \nabla z).
\]
The residual term at the right hand side is unfortunately an antidiffusive term that makes
the associated problem ill-posed. Remark that the antidiffusion matrix $A=-\bu\otimes\bu$
is negative semi-definite with a kernel $\ker(A)=\text{span}(\bu^\bot)$.
Here again, one can observe an anisotropic behaviour according to the
direction of advection. Now because in our context~$z$ is a nonsmooth
function, actually higher-order terms in the Taylor expansion are also important
(consider a scaling).
Actually the true key word here is numerical stability. One has to use numerically stable time advance schemes in order to not produce spurious interface instabilities. We will see below
in numerical experiments that even if spatial discretization in done properly, it
will lead to linear instabilities evolving into nonlinear zigzag-shaped modes if 
the explicit Euler scheme is used, whereas second-order Runge-Kutta RK2 integrators
eliminate these spurious instabilities.  
\section{Requirements for the interface capturing scheme}
%
From the above numerical study and numerical analysis, we understand that
accurate interface capturing methods must satisfy a set of requirements.
We have identified the four following ``constraints'':
\begin{enumerate}
\item The solver has to be overcompressive in the normal direction to the interface
for low-diffusive properties;
\item the solver should be second-order accurate in directions tangent to the interface;
\item the gradient limiting process has to preserve the (unlimited) gradient direction $\nabla z$ (normal direction) for accuracy purposes;
\item high-order explicit time integrators having a reasonable stability region are required.
\end{enumerate}
To fulfil the above requirements, our choice moves towards multidimensional limiting process (MLP) strategies (\cite{Yoon2009,Park2011}) that are generalizations of the limiters 
to the multidimensional case, thus allowing to keep control of the gradient direction.
For time integrators we will simply use second-order Runge-Kutta RK2 schemes.
\section{Multidimensional limiting process}
%
Multidimensional limiting process or MLP has been introduced by
different authors \cite{Yoon2009,Park2011} in order to provide an improved
accuracy for multidimensional problems, especially for high-speed computational fluid
dynamics. It is a natural extension of the one-dimensional slope-limiting 
process that takes into account the local neighbouring information for both
gradient reconstruction and limitation. MLP can be formulated on 
general unstructured grids. In our paper, we will restrict to Cartesian
grids even if the use of unstructured grids is possible.
One of the difficulties in the multidimensional case is the definition
of limitation criteria. Before MLP, older concepts like Local Extremum Diminishing (LED) proposed by Kuzmin and Turek \cite{KuzminTurek,Kuzmin} are a substitute to the one-dimensional Total Variation Diminishing (TVD) tool. The idea is to limit a
local reconstruction in order not to create new extrema, allowing for a $L^\infty$
diminishing property. \medskip

\begin{figure}[ht] \label{fig:1}
\begin{center}
\includegraphics[height=0.15\textheight]{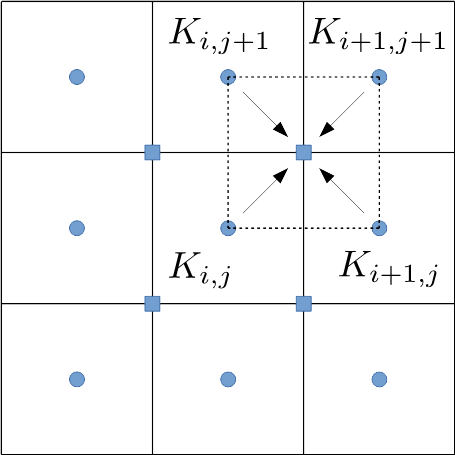}\quad\quad\quad
\includegraphics[height=0.15\textheight]{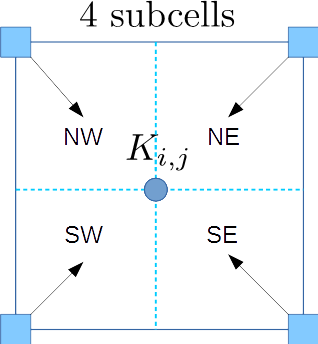}
\caption{Geometric elements for multidimensional reconstruction and limiting process.}
\label{fig:mlp}
\end{center}
\end{figure}

Here we decide to use MLP as a compressive multidimensional gradient 
reconstruction algorithm for interface capturing. Let us consider a
two-dimensional Cartesian grid made of cells named $K_{i,j}$ indexed by $(i,j)$.
The general process is the following: 
\begin{enumerate}
\item First estimate the local cell discrete gradient $(\nabla^h z)_{i,j}$. This can be performed easily
by means of approximation formulas or quadrature formulas in the case of finite volume
approximation.
\item Consider piecewise-linear local approximations in the form:
\[
\mathscr{I}^h z(\bx) = z_{i,j} + \phi_{i,j}\, (\nabla^h z)_{i,j}(\bx-\bx_{i,j}).
\]
The coefficient $\phi_{i,j}$ will be a scalar gradient limiting factor.
\item Limit the slope in order not to create new local extrema at the cell corners,
following a LED criterion~\cite{KuzminTurek}. Actually, for interface capturing we try to be as sharp as possible but without creating new local extrema, leading to a natural
extension of the overbee slope limiter.
\item Reconstruct a piecewise constant sub-square solution: the piecewise linear local solution is projected onto a piecewise constant subcell solution over the four natural corner subsquares of each cell. This projection allows us to easily compute upwind numerical fluxes.
\item Finally compute the advective fluxes at the edges.
\end{enumerate}
The involved geometric elements for performing the process are summarized on figure~\ref{fig:mlp}.
Now we give mathematical and technical details on both gradient prediction and limitation.

\subsection{Gradient reconstruction predictor step} 
%
There are probably numerous ways to determine accurate gradients. In this work,
we have chosen a finite volume approach to approximate the gradient from 
nearest neighbour cell information:
\begin{eqnarray*}
	(\nabla z)_{ij} &\approx & \frac{1}{|K_{i,j}|}\int_{K_{i,j}} \nabla z(x,y)dx\, dy \\ [1.1ex]
	                &=& \frac{1}{|K_{i,j}|}\int_{\partial K_{i,j}} z\, \bm{\nu}\, d\sigma \\ [1.1ex]
									&=& \frac{1}{|K_{i,j}|} \sum_{A\subset\partial K_{i,j}}\ \int_{A} z\, \bm{\nu}\, d\sigma.
	\end{eqnarray*}
We approximate each edge integral by second order accurate Simpson's quadrature formula, i.e. for example for edge~$A_{i+1/2,j}$:
	\[
	\int_{A_{i+1/2,j}} z\, dy \approx \frac{h}{6}\, z_{i+1/2,j+1} + \frac{2h}{3}\, z_{i+1/2,j}
	+\, \frac{h}{6} z_{i+1/2,j-1}
	\]
	where
	\[
	z_{i+1/2,j} = \frac{z_{i,j} + z_{i+1,j}}{2}.
	\]
	Summing up all the edge contributions, we get the 8-point scheme:
	\begin{equation}
	(\nabla^h z)_{i,j} =
	\frac{1}{h}\, 
	\left(\begin{array}{c}
	\frac{1}{12}(z_{i+1,j+1}-z_{i-1,j+1}) + \frac{1}{3}(z_{i+1,j}-z_{i-1,j})
	+ \frac{1}{12}(z_{i+1,j-1}-z_{i-1,j-1}) \\ [1.1ex]
	\frac{1}{12}(z_{i+1,j+1}-z_{i+1,j-1}) + \frac{1}{3}(z_{i,j+1}-z_{i,j-1})
	+ \frac{1}{12}(z_{i-1,j+1}-z_{i-1,j-1})
	\end{array}\right).
	\label{eq:gradhz}
	\end{equation}
	In particular, the quadrature formula is exact for linear functions.
\subsection{Gradient limitation correction step}
%
Now the have to limit $(\nabla^h z)_{i,j}$. For each cell $K_{i,j}$, consider
\[
\mathscr{I}^h z(\bx) = z_{i,j} + \phi_{i,j}\, (\nabla^h z)_{i,j}(\bx-\bx_{i,j}),
\quad \phi_{i,j}\geq 0.
\]
We are going to limit the gradients in order not to create new extrema at cell
corners. So, we need extrapolated corner values:
\[
\hat z_{i+1/2,j+1/2} = 
z_{i,j} + (\nabla^h z)_{i,j}(\bx_{i+1/2,j+1/2}-\bx_{i,j}).
\]
We determine local extremum values for each corner from the four
neighbor values. For example, at $\bx_{i+1/2,j+1/2}$, we compute
\[
\bar z_{i+1/2,j+1/2} = \max(z_{i,j},\, z_{i+1,j},\, z_{i,j+1},\, z_{i+1,j+1}),
\]
\[
\underline z_{i+1/2,j+1/2} = \min(z_{i,j},\, z_{i+1,j},\, z_{i,j+1},\, z_{i+1,j+1}).
\]
We ask to fulfil
\[
\underline z_{i+1/2,j+1/2} \leq \mathscr{I}^h z(\bx_{i+1/2,j+1/2}) \leq \bar z_{i+1/2,j+1/2},
\]
We then find a value $\phi_{i+1/2,j+1/2}$ computed by
\begin{eqnarray*}
\phi_{i+1/2,j+1/2} = 
&& \min\left(\beta, \frac{\bar z_{i+1/2,j+1/2}-z_{i,j}}{\hat z_{i+1/2,j+1/2}-z_{i,j}} \right) 1_{\hat z_{i+1/2,j+1/2}>z_{i,j}} \\ [1.1ex]
&& \hspace{1cm}+ 
\min\left(\beta, \frac{\underline z_{i+1/2,j+1/2}-z_{i,j}}{\hat z_{i+1/2,j+1/2}-z_{i,j}} \right) 1_{\hat z_{i+1/2,j+1/2}<z_{i,j}}
\end{eqnarray*}
for some parameter $\beta>0$ close to 1. The choice $\beta=1$ return a
second-order accurate reconstruction whereas a greater value ($\beta=2$ for example) leads to a compressive reconstruction.
We need to repeat the process for each corner of the cell $K_{i,j}$. Finally
we retain the value of the limiting factor $\phi_{i,j}$ as the minimum
of the four corner limiting factors:
\[
\phi_{i,j} = \min(\phi_{i+1/2,j+1/2},\ \phi_{i+1/2,j-1/2},\ \phi_{i-1/2,j+1/2},\ \phi_{i-1/2,j-1/2}).
\]
%
\section{Numerical experiments}
%
The present section is intended to evaluate numerical stability, accuracy and low-diffusive property of the (MLP+RK2) interface capturing strategy.
		\subsection{Uniform advection in the first diagonal direction to the mesh}
\begin{figure}[ht]
		\begin{center}
		\begin{subfigure}{0.45\textwidth}
		\includegraphics[width=\linewidth]{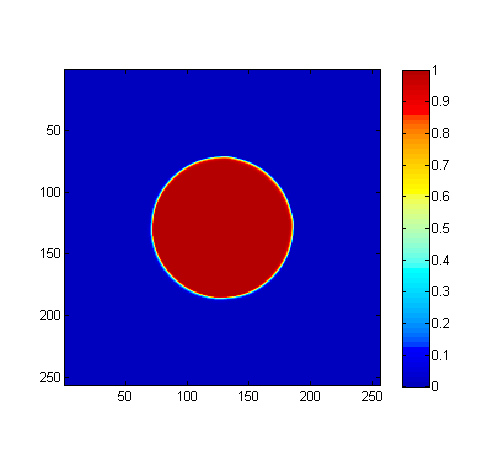}
		\caption{Colored representation of the solution $z$.}
		\end{subfigure}
		\begin{subfigure}{0.45\textwidth}
		\includegraphics[width=\linewidth]{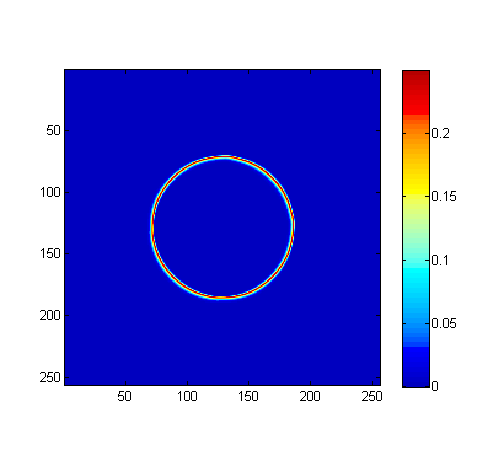}
		\caption{Colored representation of  $z(1-z)$.}
		\end{subfigure}
		\end{center}
		\caption{Disk advected into a uniform vector field $\bu=(1,1)$ in the domain
		$[-1,1]^2$ (periodic boundary conditions). 
		Discrete solution at final time $t=10$, cartesian grid $256^2$ using the MLP+RK2 strategy.} \label{fig:mlp1}
		\end{figure}

\begin{figure}[ht]
		\begin{center}
		\begin{subfigure}{0.45\textwidth}
		\begin{center}
		\includegraphics[width=\linewidth]{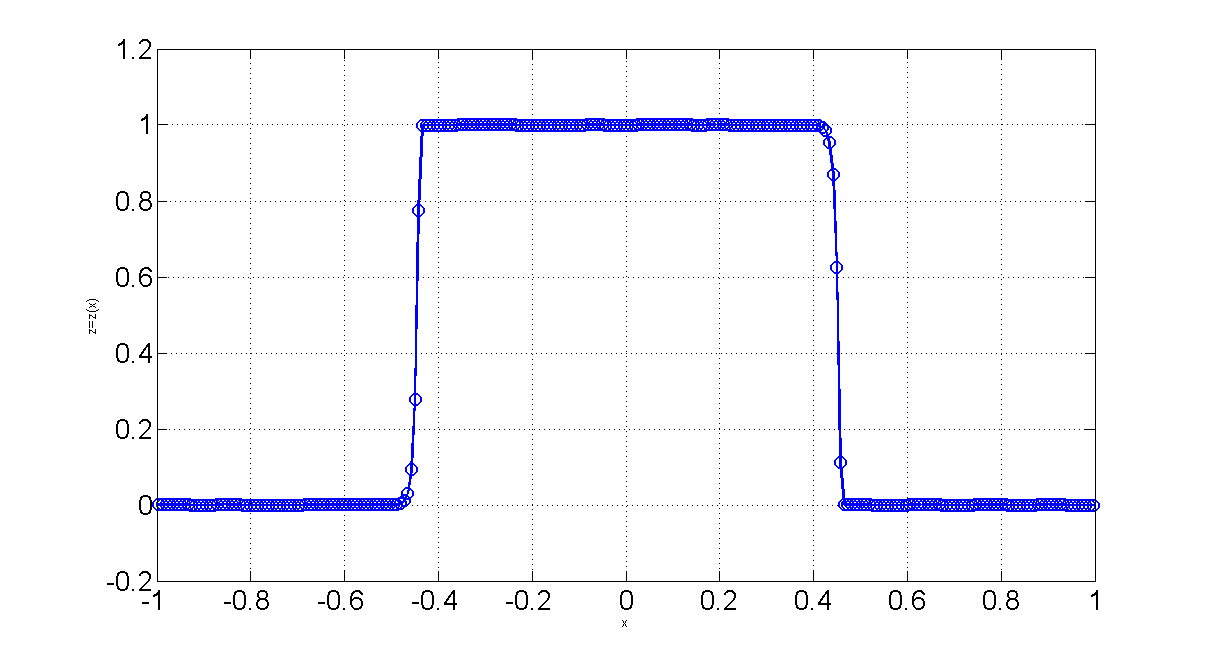}
		\caption{Horizontal cut plane $(y=0)$.}
		\end{center}
		\end{subfigure}
		\begin{subfigure}{0.45\textwidth}
		\begin{center}
		\includegraphics[width=\linewidth]{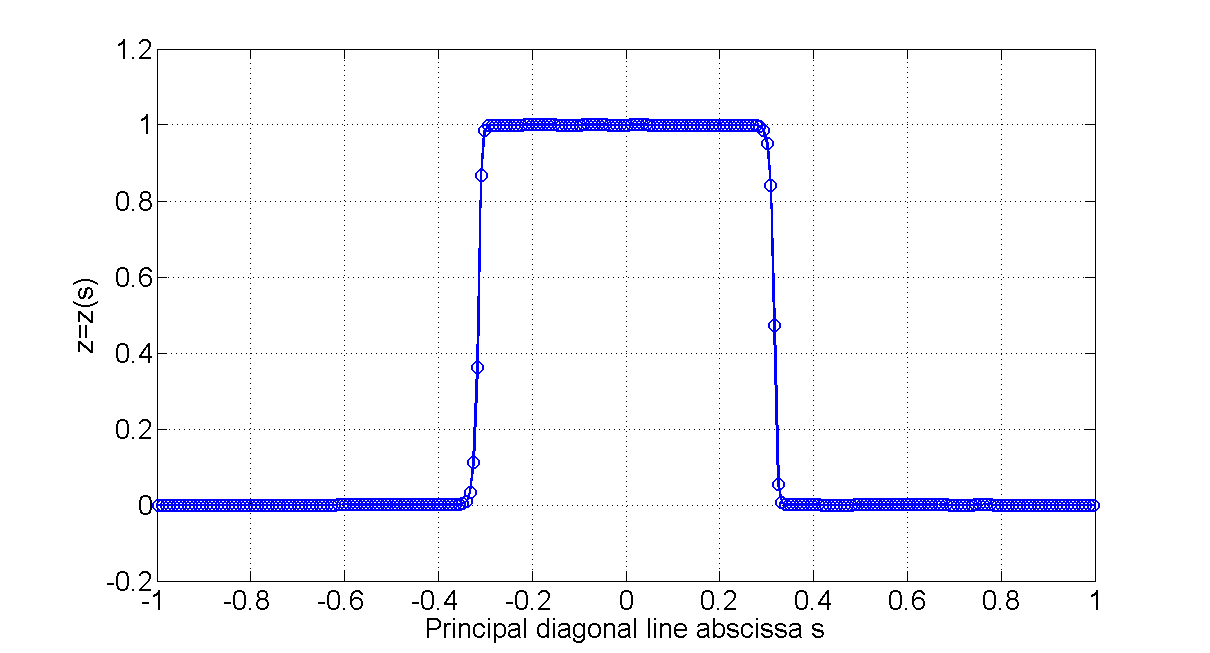}
		\caption{principal diagonal cut plane.}
		\end{center}
		\end{subfigure}
		\end{center}
		\caption{Plots of $z$ profiles for different cut planes.}\label{fig:mlp2}
		\end{figure}

		%
		Here we take again the advection case of uniform advection in the direction diagonal to the grid as introduced in section~\ref{sec:22}. After several
		cycles of advection, on figure~\ref{fig:mlp1} we can observe that the artefacts completely dispappear with the (MLP+RK2) strategy. The price to pay is a stronger
		numerical diffusion, but the interface smearing stays reasonable. 
		On figure~\ref{fig:mlp2} the profile
		of $z$ on the cut plane $y=0$  and the cut plane $y=x$ respectively show
		the low-diffusive capture of interface discontinuities.	
		\subsection{Advection of a disk into a pure rotation vector field}
		%
		We have also tested the case of rigid rotation of a disk from section~\ref{sec:22}. This time, applying the (MLP+RK2) strategy leads to an accurate transport of the 
		disk, free from any zigzag instabilities. The profile
		on an horizontal cut plane again shows sharp discontinuities after one
		disk revolution.
		%
		%
		\subsection{Experimental error measurements on the stationary oblique discontinuity problem}
\begin{figure}[ht]
		\begin{subfigure}{0.32\textwidth}
		\includegraphics[width=\linewidth]{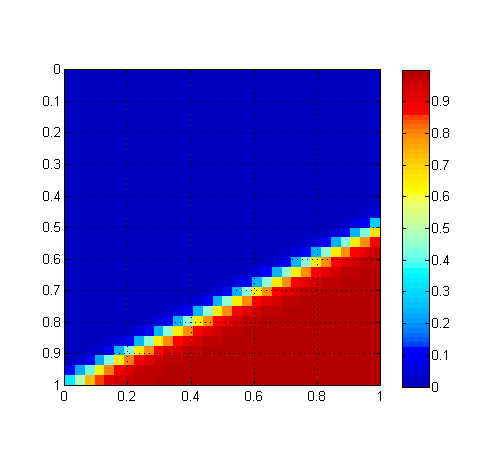}
		\caption{Grid $32\times 32$}
		\end{subfigure}
		\begin{subfigure}{0.32\textwidth}
		\includegraphics[width=\linewidth]{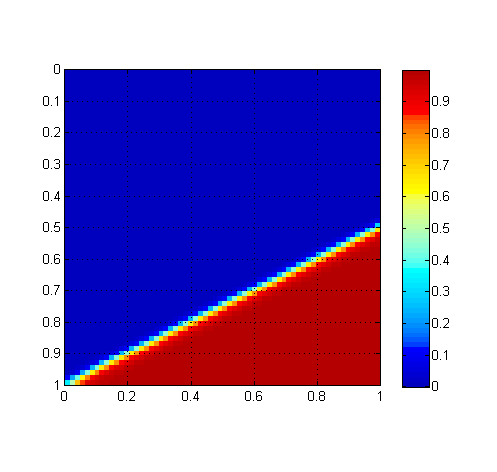}
		\caption{Grid $64\times 64$}
		\end{subfigure}
		\begin{subfigure}{0.32\textwidth}
		\includegraphics[width=\linewidth]{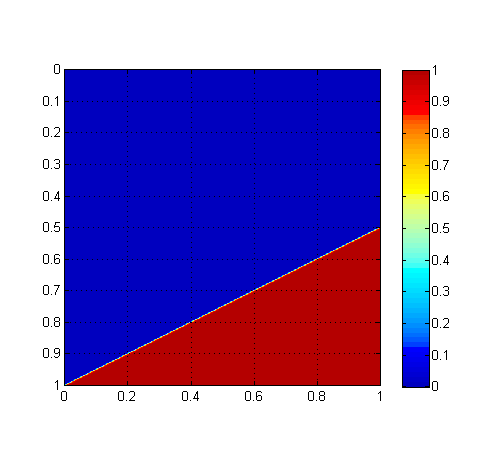}
		\caption{Grid $512\times 512$}
		\end{subfigure}
		\caption{Convergence analysis on the oblique discontinuity stationary problem}
		\label{fig:mlp6}
		\end{figure}

\begin{figure}[ht]
		\begin{subfigure}{0.48\textwidth}
		\includegraphics[width=\linewidth]{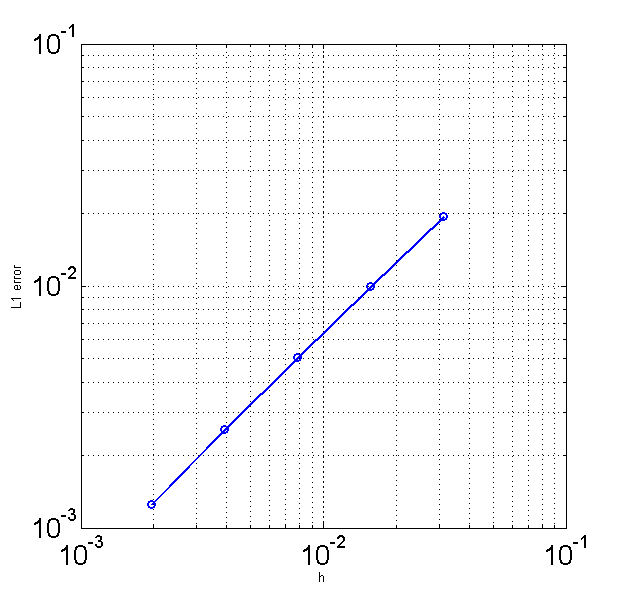}
		\caption{Convergence analysis in $L_1$ norm\\ (log scale). Slope = 0.9861
		(almost 1).}
		\end{subfigure}
		\begin{subfigure}{0.48\textwidth}
		\includegraphics[width=\linewidth]{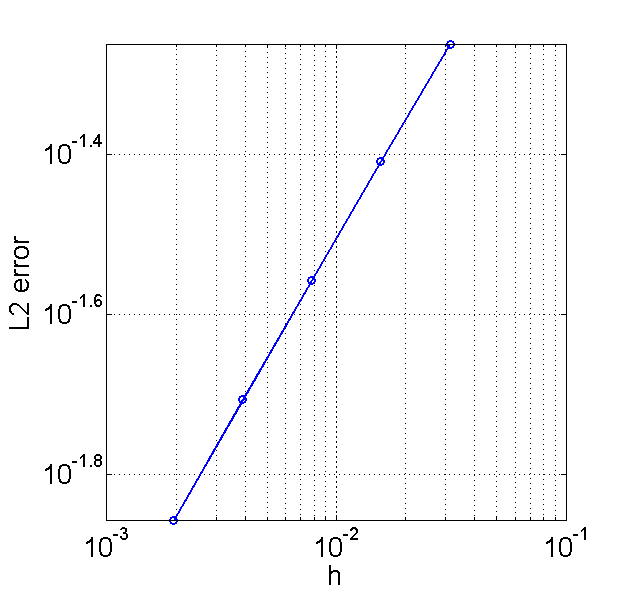}
		\caption{Convergence analysis in $L_2$ norm\\ (log scale). Slope = 0.494
		(almost $1/2$).}
		\end{subfigure}
		\caption{Convergence rates in $L_1$ and $L_2$ norms. Note that the convergence rate of the quantity $E(h) = \left|\int_\Omega (z^h-z^\infty)\, d\bx\right|$ is 2.}
		\label{fig:mlp7}
		\end{figure}

		%
		To complete the experiments, we take again the third case of steady oblique discontinuity presented in section~\ref{sec:22} and perform a convergence
		analysis. On figure~\ref{fig:mlp6}, we plot the discrete steady field obtained
		with MLP+RK2 for different grid refinement: $32\times 32$, $64\times 64$
		and $512\times 512$ respectively. More interesting is the measured convergence
		rates shown on figure~\ref{fig:mlp7}. We find a numerical convergence 
		of~$0.9861$ (close to 1) for the $L^1$-norm and $0.494$ (close to 1/2) for the
		$L^2$-norm, showing the accuracy level of the method. 
		Remark also that the measured global conservation error convergence
		\[
		E(h) = \left|\int_\Omega (z^h-z^\infty)\, d\bx\right|
		\]
		is not zero exactly because of second-order errors of boundary outstream fluxes,
		but anyway it is second-order accurate. 
		\subsection{Zalesak's disk reference case}
\begin{figure}[ht]
		\begin{center}
		\begin{subfigure}{0.45\textwidth}
		\includegraphics[width=\linewidth]{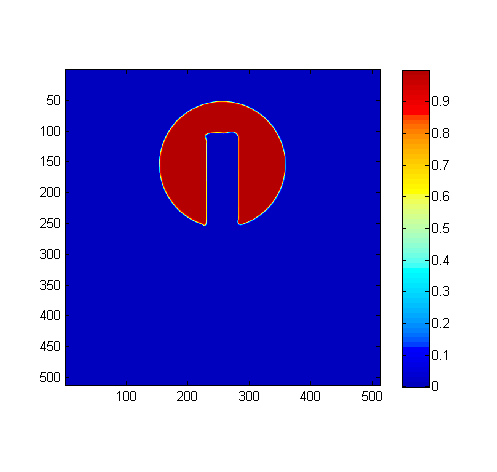}
		\caption{Colored representation of the solution $z$ after one revolution.}
		\end{subfigure}
		\begin{subfigure}{0.45\textwidth}
		\includegraphics[width=\linewidth]{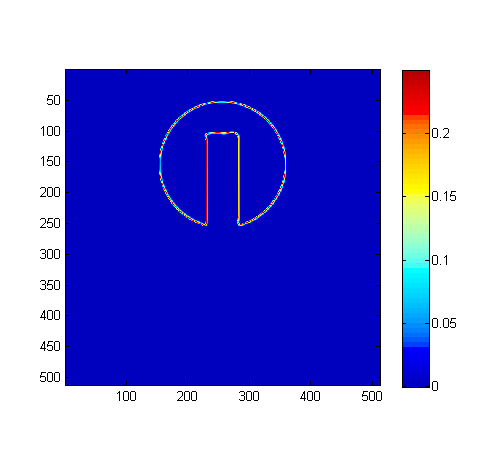}
		\caption{Colored representation of  $z(1-z)$ after one revolution.}
		\end{subfigure}
		\end{center}
		\caption{Zalesak's disk reference case, grid $512\times 512$.} \label{za:1}
		\end{figure}

\begin{figure}[ht]
		\begin{center}
		\includegraphics[height=0.2\textheight]{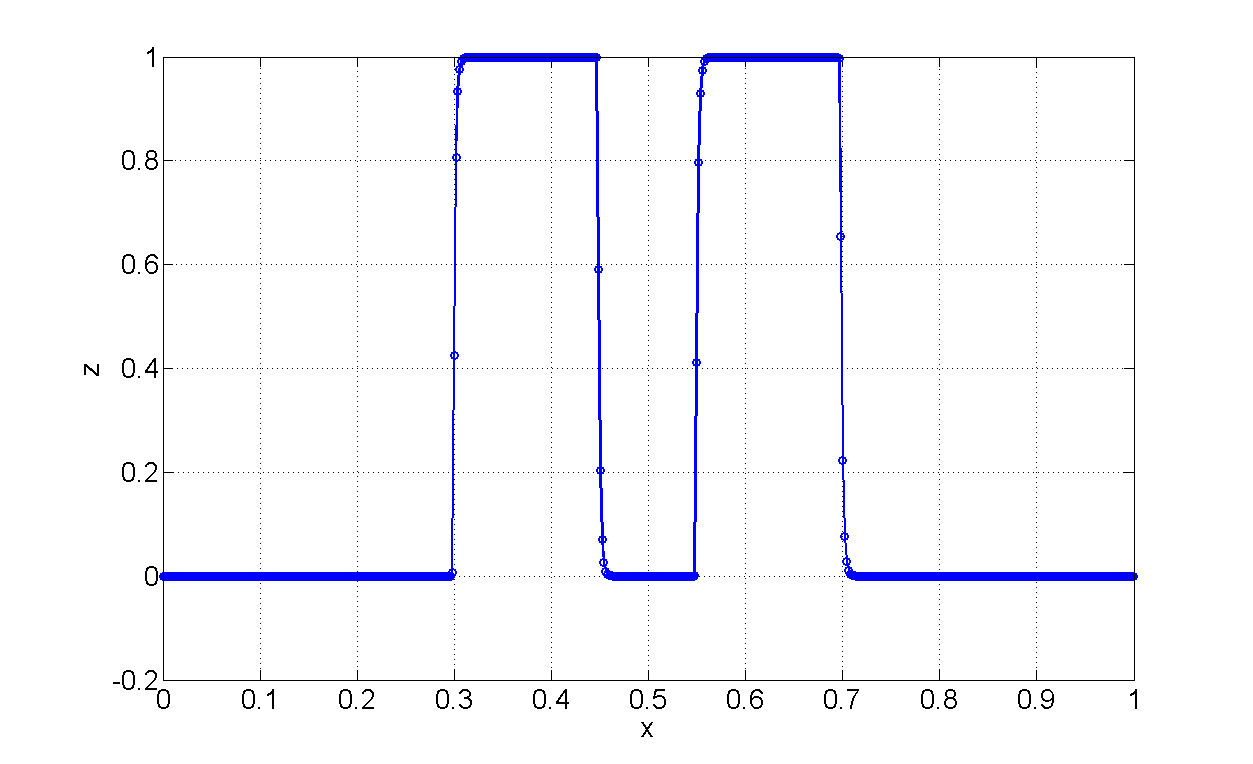}
		\end{center}
		\caption{Zalesak's disk: $z$ profile at the cut plane $y=7/10$ after one revolution.} \label{za:2}
		\end{figure}

	  %
		We consider here the classical test case of rigid rotation of Zalesak's disk
		in a constant rotating velocity field \cite{Zalesak}: $\bu(x,y)=(1/2-y, x-1/2)$
		on the domain $\Omega=(0,1)^2$. The initial data $z^0$ is associated to
		a slotted circle centered at $(1/2,7/10)$ with a radius $r=1/5$, the slot
		depth is $3/10$ and the width is equal to $1/10$. We use a cartesian
		mesh composed of $512\times 512$ points. On figure~\ref{za:1}, the $z$
		field is plotted after one revolution ($t=2\pi$). One can observe the rather good
		accuracy of the discrete solution. Some corner effects can be observed, mainly
		due to the fact that the solution is not differentiable at corners, leading
		to discrete gradient inaccuracies. On figure~\ref{za:2} the $z$ profile along the line $y=7/10$ showing sharp discontinuities and the low-diffusive property using MLP. 
		\subsection{Reference Kothe-Rider forward-backward advection and stretching case}
\begin{figure}[ht]
		\begin{subfigure}{0.5\textwidth}
		\includegraphics[width=0.99\linewidth]{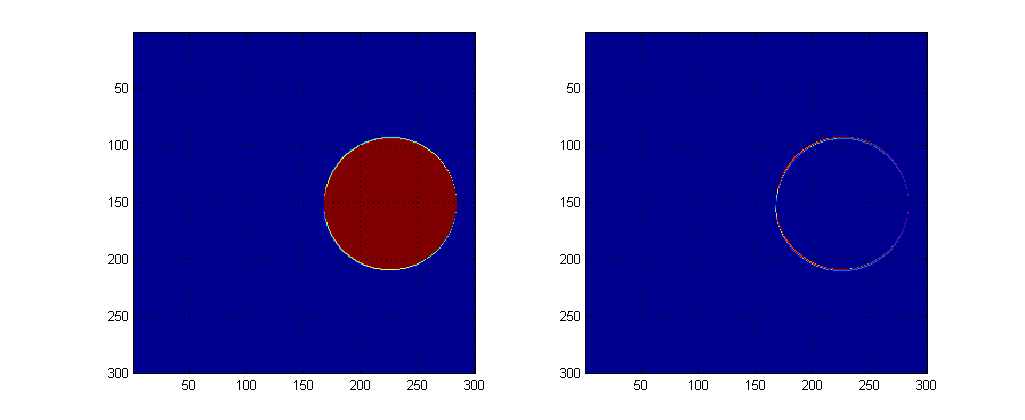}
		\caption{At initial time $t=0$}
		\end{subfigure}
		\begin{subfigure}{0.5\textwidth}
		\includegraphics[width=0.99\linewidth]{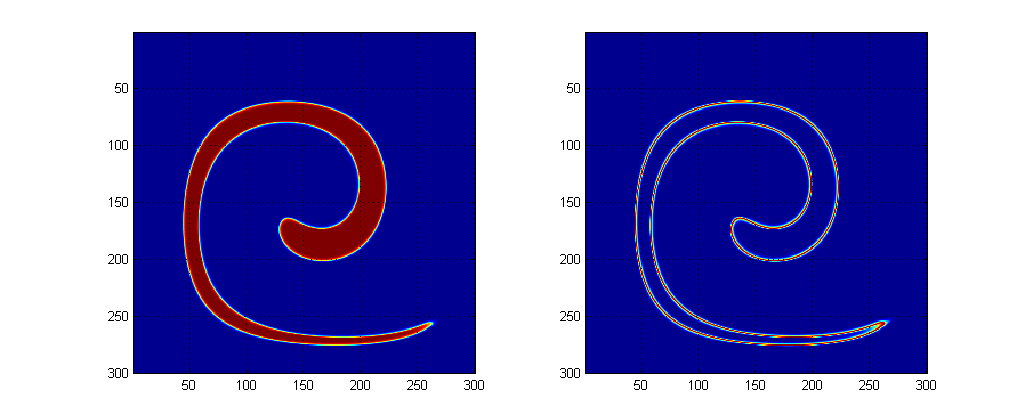}
		\caption{At time $t=3$}
		\end{subfigure}
		\begin{subfigure}{0.5\textwidth}
		\includegraphics[width=0.99\linewidth]{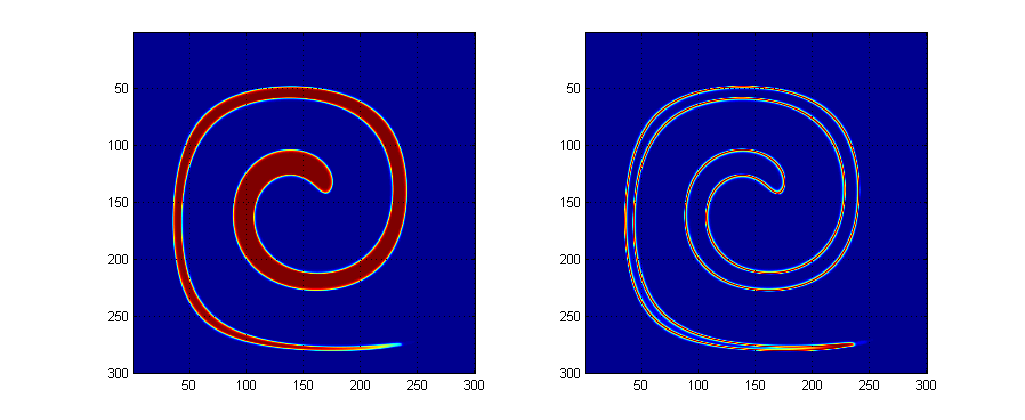}
		\caption{At time $t=6$}
		\end{subfigure}
		\begin{subfigure}{0.5\textwidth}
		\includegraphics[width=0.99\linewidth]{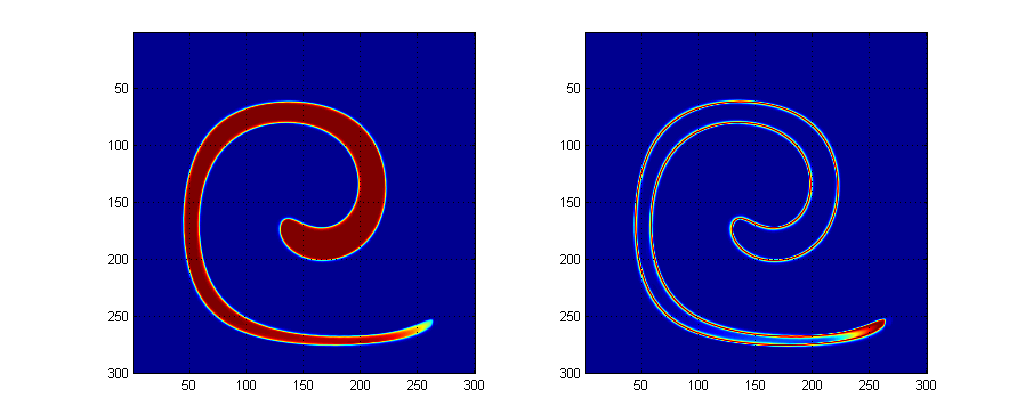}
		\caption{At time $t=9$}
		\end{subfigure}
		\begin{subfigure}{0.5\textwidth}
		\includegraphics[width=0.99\linewidth]{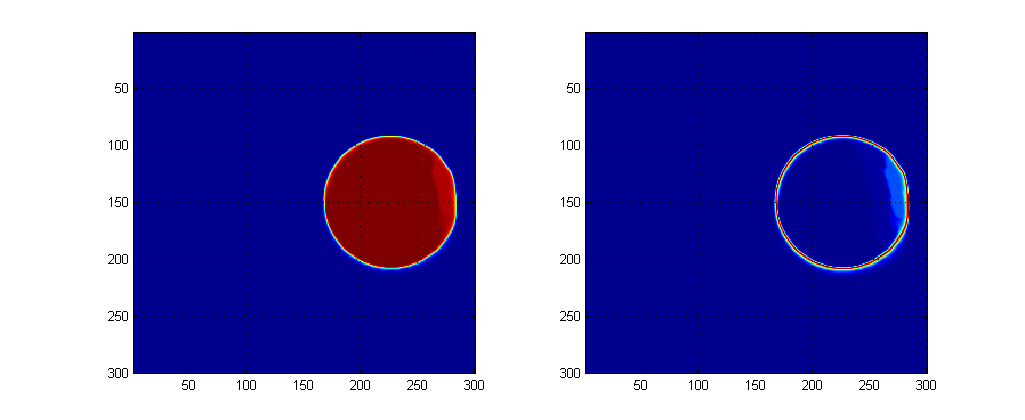}
		\caption{At final time $t=12$}
		\end{subfigure}
		\begin{subfigure}{0.5\textwidth}
		\includegraphics[width=0.99\linewidth]{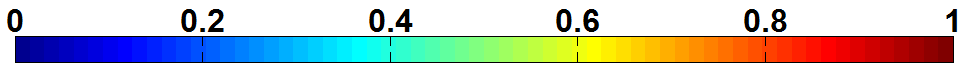}
		\end{subfigure}
		\caption{Validating the MLP+RK2 strategy on the Kothe-Rider advection case, cartesian mesh grid $300^2$.} \label{fig:kr1}
		\end{figure}

\begin{figure}[ht]
		\begin{subfigure}{0.5\textwidth}
		\includegraphics[width=0.99\linewidth]{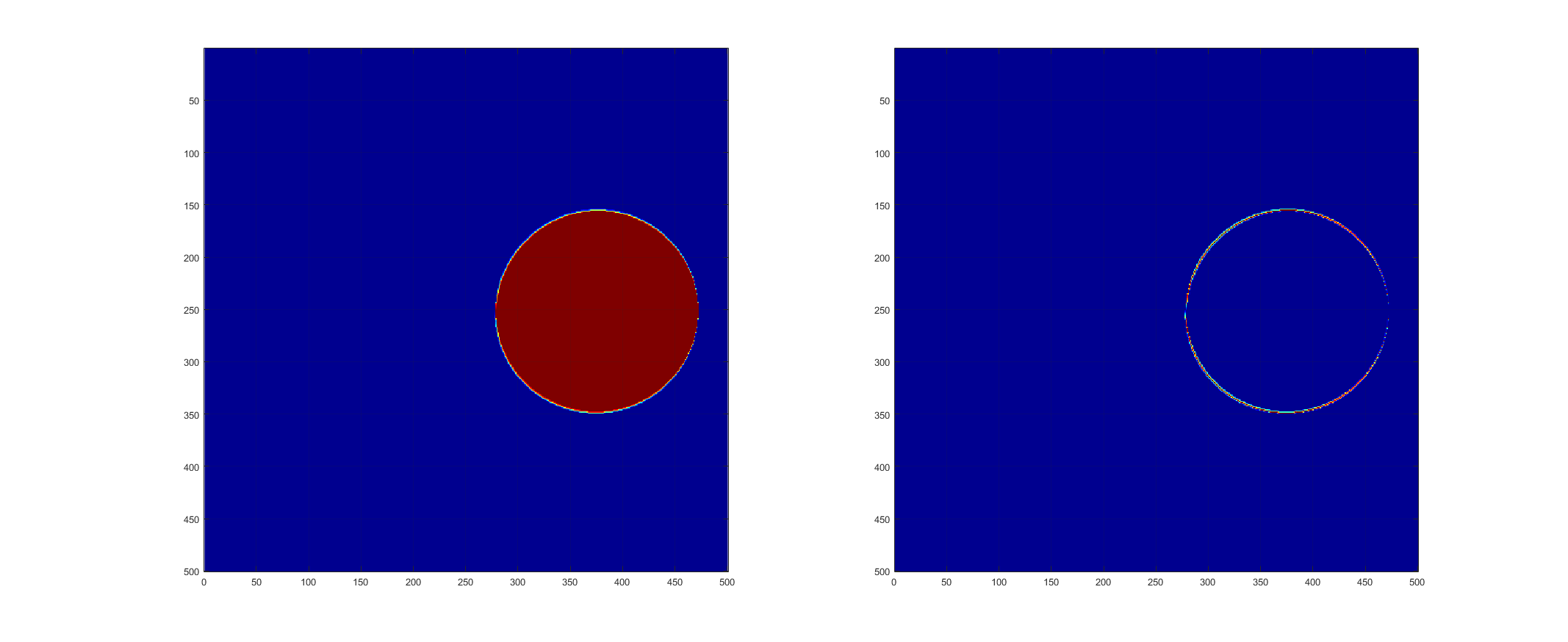}
		\caption{At initial time $t=0$}
		\end{subfigure}
		\begin{subfigure}{0.5\textwidth}
		\includegraphics[width=0.99\linewidth]{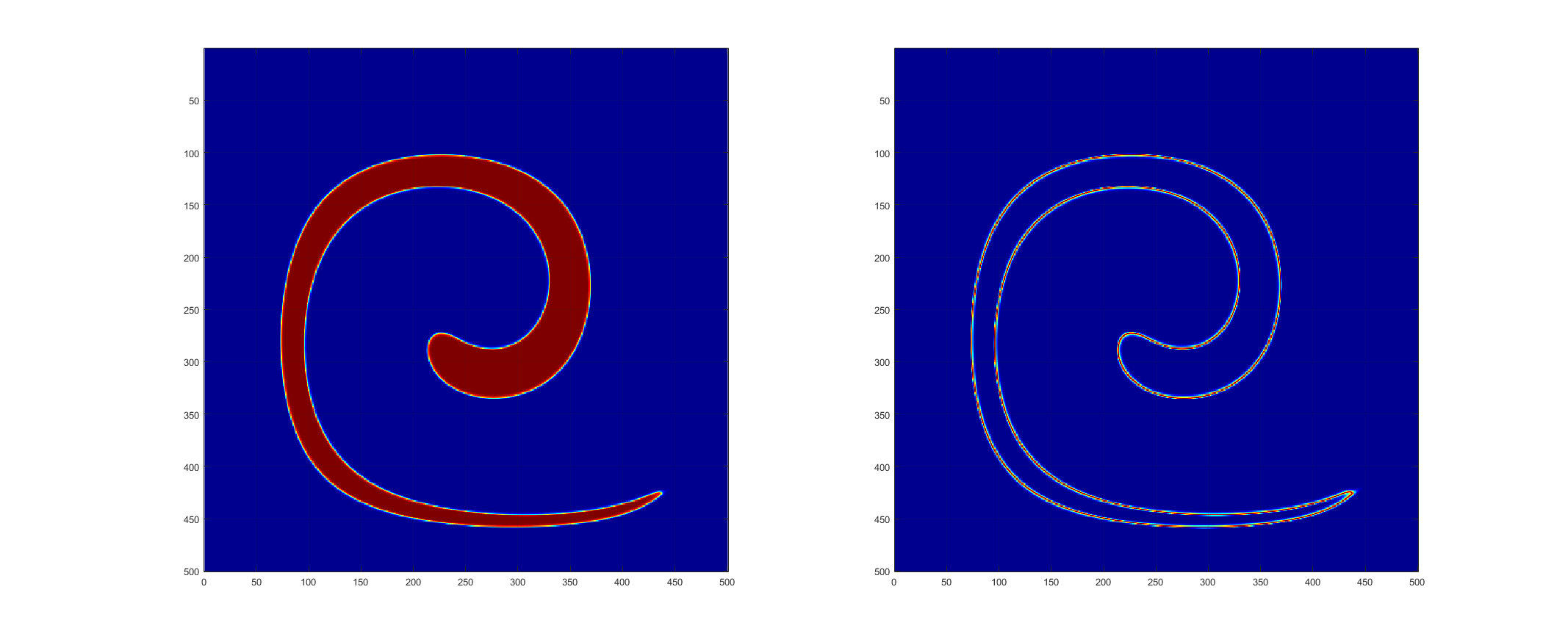}
		\caption{At time $t=3$}
		\end{subfigure}
		\begin{subfigure}{0.5\textwidth}
		\includegraphics[width=0.99\linewidth]{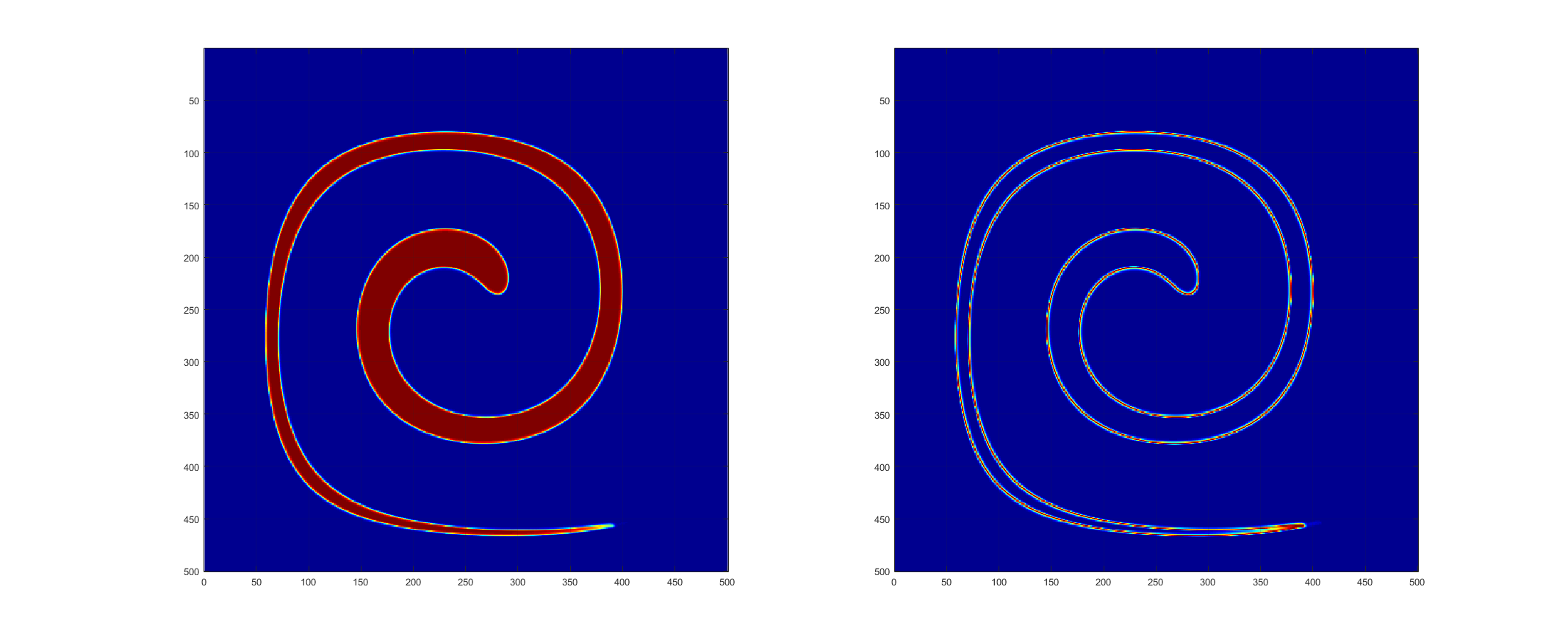}
		\caption{At time $t=6$}
		\end{subfigure}
		\begin{subfigure}{0.5\textwidth}
		\includegraphics[width=0.99\linewidth]{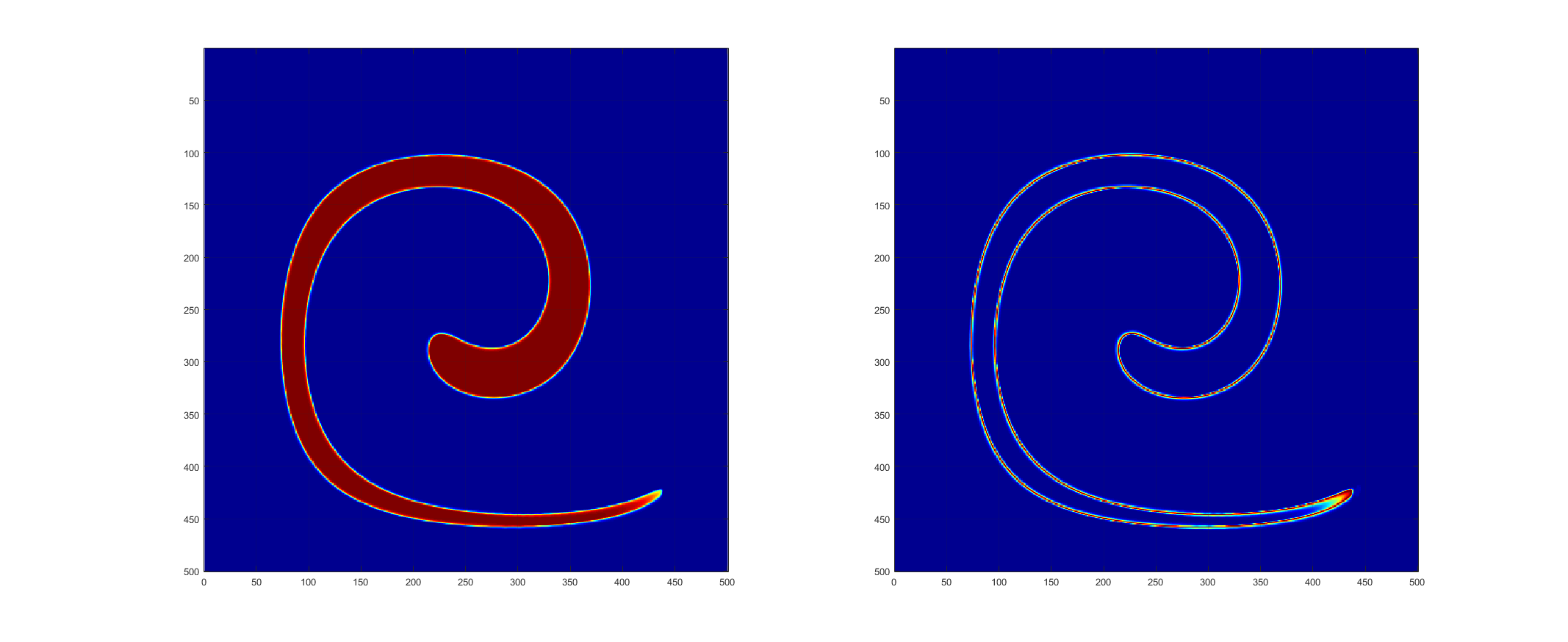}
		\caption{At time $t=9$}
		\end{subfigure}
		\begin{subfigure}{0.5\textwidth}
		\includegraphics[width=0.99\linewidth]{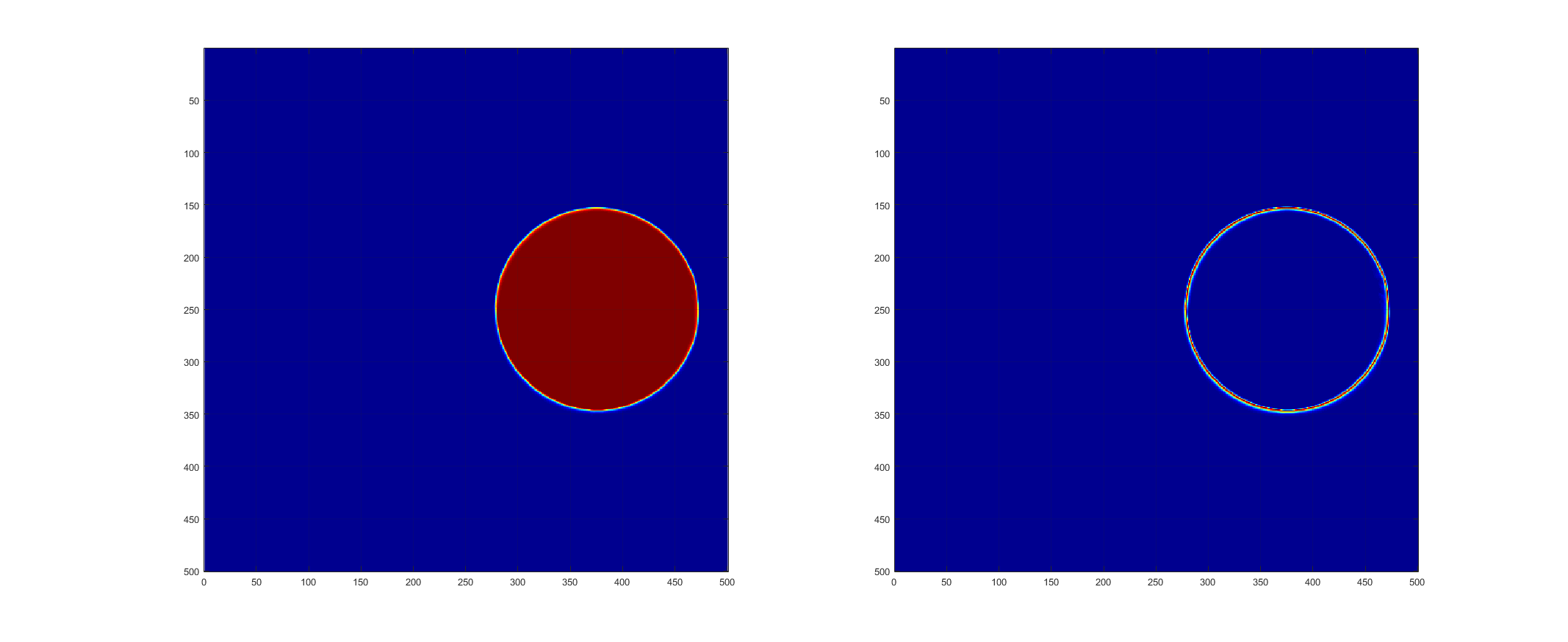}
		\caption{At final time $t=12$}
		\end{subfigure}
		\begin{subfigure}{0.5\textwidth}
		\includegraphics[width=0.99\linewidth]{hcolorbar}
		\end{subfigure}
		\caption{Validating the MLP+RK2 strategy on the Kothe-Rider advection case, cartesian mesh grid $500^2$.} \label{fig:kr2}
		\end{figure}

		%
		The reference case proposed to Rider and Kothe~\cite{Rider} is the deformation
		of a disk into a rotating and stretching velocity field. A forward-then-backward
		velocity field allows to come back to the initial condition (reversible 
		process) and thus to assess the accuracy of the approach by measuring the 
		deviation between the final time solution and the initial one. 
		We performed two tests with two different grid levels, the first one with
		a grid composed of $300\times 300$ cells (figure~\ref{fig:kr1}) and the second
		one with a grid $500\times 500$ (figure~\ref{fig:kr2}). Using the coarsest grid, one can observe
		a smearing regions in the final solution, mainly to the numerical diffusion
		produced by the stretching process, but the initial disk shape is rather
		preserved. For the $500^2$-grid, the discrete solution at final
		time is very satisfactory with a preserved disk shape, showing the accuracy of the approach. The low-diffusive interface capturing process has been able to capture
		the stretching effect without any artefacts or instabilities.
		%
		%
%
%
%
%
\section{Extending to compressible multimaterial flows}
		%
		%
		In this section we would like to give ideas and highlights on how the method can be extended 
		to compressible multifluid/multimaterial hydrodynamic flows in the presence of several
		immiscible fluids.
		Even if fluids are assumed be immiscible, we use volume averaged-like balance equations
		because of finite volume averages that may create what we call ``mixed cells'' when
		more than one fluid in present in a cell. 
		Of course, for the continuous problem, volume fractions should be either~0 or~1.
		Consider the mass, momentum and total
		density energy conservation equations for a system of inviscid fluids
		\begin{eqnarray*}
		&& \partial_t (\alpha_\ell \rho_\ell) + \nabla \cdot (\alpha_\ell \rho_\ell \bu) = 0, 
		\quad \ell=1,...,N,\\ [1.1ex]
		&& \partial_t (\rho \bu) + \nabla\cdot(\rho \bu\otimes\bu) + \nabla p = 0,\\ [1.1ex]
		&& \partial_t (\rho E) + \nabla\cdot (\rho E\bu) + \nabla\cdot (p\bu) = 0,
		\end{eqnarray*}
		where $\rho_\ell$, $\ell=1,...,N$ denote the partial densities of each fluid, the
		variables $\alpha_\ell\in [0,1]$ are the volume fractions of each fluid, $p$ denote the 
		pressure of the fluid (we assume local mechanical equilibrium), 
		$\rho=\sum_{\ell=1}^N \alpha_\ell \rho_\ell$ is the volume-averaged density and $\rho E$ is the 
		``mixture'' total density energy, sum of both kinetic and internal density energies:
		\[
		\rho E = \frac{1}{2}\rho |\bu|^2 + \sum_{\ell=1}^N \alpha_\ell \rho_\ell e_\ell.
		\]
		To this system we add the volume compatibility relation
		\[
		\sum_{\ell=1}^N \alpha_\ell = 1
		\]
		and equations of states (EOS) of each fluid, linking both density and internal energy to the fluid pressure and temperature:
		\[
		\rho_\ell=\rho_\ell(p_\ell,T_\ell), \quad\quad
		e_\ell = e_\ell(p_\ell, T_\ell).
		\]
		To close the system, we will here assume the simplest closure of local
		temperature equilibrium and local pressure equilibrium, i.e.
		\[
		T_1=...=T_N=T,\quad \quad p_1=...=p_N=p.
		\]
		For simplicity, we will assume here a system of perfect gases, where 
		\[
		p_\ell = \rho_\ell \frac{R}{M_\ell} T_\ell, \quad e_\ell = c_{v;\ell} T_\ell,
		\]
		with $R$ the universal constant of perfect gases, $M_\ell$ the molar mass of
		fluid $\ell$, $c_{v;\ell}>0$ the (constant) specific heat at constant volume. We have also
		the relation
		\[
		\gamma_\ell-1 = \frac{c_{v;\ell}R}{M_\ell},
		\]
		where $\gamma_\ell>1$ is the (constant) specific heat ratio so that 
		\[
		p_\ell = (\gamma_\ell -1) \rho_\ell e_\ell.
		\]
		It is known that the resulting system of partial differential
		equations is hyperbolic on the admissible state space of
		positive densities and positive temperatures. For more general equations of state like
		stiffened gases and well-posedness issues, see~\cite{Flatten} for example.
		
		From the conservative variables $m_\ell := \alpha_\ell \rho_\ell$, $\rho \bu$ and $\rho E$
		one can compute the primitive variables following the calculation sequence
		\begin{eqnarray*}
		&& \rho= \sum_{\ell=1}^N m_\ell, \quad \bu=\frac{\rho \bu}{\rho},\quad 
		T = \frac{\rho E - 1/2 \rho |\bu|^2}{\sum_{\ell=1}^N c_{v;\ell} m_\ell}, \\ [1.1ex]
		&& p = \sum_{\ell=1}^N \alpha_\ell \ p = \sum_{\ell=1}^N (\gamma_\ell-1) m_\ell c_{v;\ell} T, \\ [1.1ex]
		&& e_\ell = c_{v;\ell} T, \\ [1.1ex]
		&& \rho_\ell = \frac{1}{\gamma_\ell-1} \frac{p}{e_\ell}, \ \quad
		\alpha_\ell = \frac{m_\ell}{\rho_\ell}, \quad
		\rho=\sum_{\ell=1}^N \alpha_\ell \rho_\ell
		\end{eqnarray*}
		without ambiguity. Rather than volumes fraction $\alpha_\ell$, one could also
		use mass fractions $z_\ell$ defined by
		\[
		z_\ell = \frac{\alpha_\ell \rho_\ell}{\rho}.
		\]
		Then mass conservation equations read
		\[
		\partial_t (\rho z_\ell) + \nabla\cdot (\rho z_\ell \bu) = 0.
		\]
		Then from the continuity equation $\partial_t\rho + \nabla\cdot (\rho\bu)=0$, one can observe
		that the variables $z_\ell$ are advected according to the fluid velocity. For smooth 
		solutions of the problem we have the equivalent transport equations
		\[
		\partial_t z_\ell + \bu\cdot\nabla z_\ell = 0.
		\]
		Thus we want to apply our interface capturing approach to the variables $z_\ell$,
		while guaranteeing that the numerical scheme is conservative on all the conservative
		variables and ensuring global numerical stability. Remark that we can pass from volume fraction variables to 
		mass fractions by the direct and inverse formulas:
		\[
		z_\ell = \frac{\alpha_\ell\rho_\ell}{\sum_{m=1}^N \alpha_m\rho_m}, \quad\quad
		\alpha_\ell = \frac{z_\ell \tau_\ell}{\sum_{m=1}^N z_m\tau_m}.
		\]
		with $\tau_m = (\rho_m)^{-1}$ as specific volumes. Of course,
		we have again the compatibility relation on the mass fractions~$\sum_{\ell=1}^N z_\ell=1$. 
		\subsection{Lagrange+remap scheme}
		%
    Our strategy of discretization follows the
		classical remapped Lagrange schemes made of two steps: i) first step is a full solution
		of the problem with a Lagrangian flow description; ii) a remap step allowing to 
		project the quantities onto the reference (Eulerian) mesh. 
		Let us assume that the Lagrange step, in which mass fractions are kept unchanged,
		is correctly solved (use for example collocated Lagrange solvers proposed
		by Despr\'es-Mazeran~\cite{Mazeran} or Maire et al.~\cite{Maire}). 
		We rather shall focus on the remap step, seen as a convective flux
		step. \medskip
		
		Rather than performing geometrical projections that involves mesh intersections, 
		we use instead a convective flux formulation that involves material fluxes through
		the edges of the Eulerian mesh. If~$\bv$ denotes the Lagrangian velocity vector field
		(usually chosen as a Lagrangian velocity field $\bu^{n+1/2}$ at middle time $t^{n+1/2}$ for accuracy
		purposes), we claim that the remapping step is nothing else but the solution of the convection
		system
		\[
		\partial_t (\rho W) + \nabla\cdot(\rho W \bv) = 0.
		\]
		with the vector of variables $W=([z_\ell]_\ell, \bu, E)$ other a time step $\Delta t^n$ (see De Vuyst et al.~\cite{DeVuyst2016}).
		 We want to design the numerical convective fluxes in order to get accuracy,
		stability and low-diffusivity of the mass fractions~$z_\ell$. Remark that, because
		\[
		\partial_t \rho + \nabla\cdot (\rho \bv) = 0
		\]
		in the remap step, actually all the variables of the $W$ vector are advected:
		\[
		\partial_t W + \bv\cdot\nabla W = 0
		\]
		and thus could satisfy a discrete local maximum principle in the step,
		providing a stability result for this step (in $L^\infty$ norm for example). \medskip
		
		In order to control the mass fractions $z_\ell$ in $L^\infty$ norm, we shall follow ideas from Larrouturou~\cite{Larrouturou}: considering a (semi-discrete)
		mass balance finite volume scheme
		\[
		\frac{d\rho_K}{dt} = -\frac{1}{|K|}\sum_{A\subset \partial K} |A| \Phi_{\rho,A}
		\]
		for a total mass flux $\Phi_{\rho,A}$  through the edge $A$, we consider
		partial mass balance schemes in the form
		\[
		\frac{d(\rho z_\ell)_K}{dt} = -\frac{1}{|K|}\sum_{A\subset \partial K} |A| 
		(z_\ell)_A \, \Phi_{\rho,A}
		\]
		with a value of mass fraction $(z_\ell)_A$ at the edge $A$, to define. 
		Proceeding like that, one can check that we have
		\[
		\frac{d (z_\ell)_K}{dt} = -\frac{1}{|K|}  \sum_{A\subset \partial K} |A| 
		\left[(z_\ell)_A-(z_\ell)_K\right] \, \frac{\Phi_{\rho,A}}{\rho_K}.
		\]
		Under CFL-like conditions of the type
		\[
		\frac{\Delta t\, |A|}{|K|}\,  \frac{|\Phi_{\rho,A}|}{\rho_K} \leq 1
		\quad\quad \forall A\in\partial K, \ \forall K,
		\]
		it is not difficult to derive explicit first-order schemes that fulfill
		a local discrete maximum principle. For that, it is natural to introduce
		upwind edge values $(z_\ell)_A$ according to the sign of the mass 
		flux~$\Phi_{\rho,A}$. Upwind values will be then denoted $(z_\ell)_A^{upw}$ in the
		sequel.
		%
		%
		\paragraph{Contact discontinuities and pressure oscillations.} As emphasized
		by many authors, conservative schemes may lead to important concentrated pressure 
		oscillations through contact discontinuities (see \cite{Saurel} for instance) for multifluid
		flows. The reason behind that is that there are incompatibilities between
		the numerical viscous profile of total density or partial densities and
		profiles of mass fractions.
		
		To fix this problem, is it important to compute edge quantities that are
		compatible with the (current local) pressure. The leading algorithm is
		the following one~:
		\begin{enumerate}
		\item For each cell $K$, compute the pressure variable $p_K$ and temperature variable $T_K$;
		\item Compute the associated partial densities 
		$(\rho_\ell)_K = \rho_\ell(p_K,T_K)$;
		\item Compute the edge mass fractions $(z_\ell)_A$ from the MLP algorithm presented in a previous section;
		\item Then deduce from the $(z_\ell)_A$ and $(\rho_\ell)_K$ the volume
		fractions~:
		\[
		(\alpha_\ell)_A = \frac{(z_\ell)_A (\tau_\ell)_K}{\sum_{m=1}^N (z_m)_A (\tau_m)_K};
		\]
		\item Compute a mean edge density $\rho_A$ as
		\[
		\rho_A = \sum_{\ell=1}^N (\alpha_\ell)_A (\rho_\ell)_K
		\]
		\item For each edge, compute the upwind edge density $\rho_A^{upw}$ from the
		extrapolated values $\rho_A$ and deduce a mass flux
		\[
		\Phi_{\rho,A} = \rho_A^{upw}\, (\bv_A\cdot\nu_A).
		\]
		\item For each edge, deduce the upwind edge volume fractions 
		$(z_\ell)_A^{upw}$ according to the sign of the mass flux $\Phi_{\rho,A}$;
		\item Integrate the semi-discrete scheme
		\[
		\frac{d(\alpha_\ell \rho_\ell)_K}{dt} = -\frac{1}{|K|}\sum_{A\subset \partial K} |A| 
		(z_\ell)_A^{upw} \, \Phi_{\rho,A}
		\]
		over a time step $\Delta t$.
		\end{enumerate}
		Of course this algorithm can be easily extended to second-order accuracy
		in space. For that, consider MUSCL reconstructions on the thermodynamic
		variables $p$ and $T$. Then for each cell $K$ we have to compute
		extrapolated values of pressure and temperature $p_A$ and $T_A$ 
		respectively at edge $A$, then compute partial densities $(\rho_\ell)_A$ as
		\[
		(\rho_\ell)_A = \rho_\ell(p_A, T_A),
		\]
		and edge volume fractions 
		\[
		(\alpha_\ell)_A = 
		\frac{(z_\ell)_A (\tau_\ell)_A}{\sum_{m=1}^N (z_m)_A (\tau_m)_A},
		\]
		then complete by the above algorithm to finish.
		%
		%
		\paragraph{Gradients compatibility.}
		%
		For more than two fluids, linear reconstructions must pay attention
		to the mass compatibility invariant
		\[
		\sum_{\ell=1}^N z_\ell = 1.
		\]
		The MLP reconstruction requires both prediction and limitation of each
		gradient per species~$\ell$. For a gradient reconstruction on the
		mass fraction $z_\ell$ par cell $K$, i.e.
		\[
		\mathscr{I}^h z_\ell(\bx) = (z_\ell)_K + (\phi_\ell)_K
		(\nabla^h z_\ell)_K\cdot(\bx-\bx_K), 
		\]
		because we want for any $\bx$ in $K$ to have
		\[
		\sum_{\ell=1}^n \mathscr{I}^h z_\ell(\bx) = 1, 
		\]
		we get the expected compatibility formula on the limited gradients
		\begin{equation}
		\sum_{\ell=1}^N (\phi_\ell)_K (\nabla^h z_\ell)_K = \bm{0}.
		\label{eq:compat}
		\end{equation}
		As a first remark, if the apply the gradient prediction formula~\eqref{eq:gradhz}
		for each fluid $\ell$, by linearity of the formula we clearly have
		\[
		\sum_{\ell=1}^N (\nabla^h z_\ell)_K = \bm{0}.
		\]
		The difficulty here is once again due to the linear procedure of limitation
		which may violate the identity~\eqref{eq:compat}. We propose the following
		algorithm: first, apply the MLP, as explained in previous sections, for each
		fluid $\ell$. We get an estimator of limiting factor for each $\ell$, 
		here denoted by~$\widehat{(\phi_\ell)}_K$, $\widehat{(\phi_\ell)}_K\geq 0$.
		
		In a second step, we have to find a procedure to more limit these factors
		in order to satisfy the identity~\eqref{eq:compat}. This can be done for example
		by the use of a linear-quadratic (LQ) optimization problem
		\begin{equation}
		\max_{(\phi_1,\phi_2,...,\phi_N)}\quad
		\frac{1}{2} \sum_{\ell=1}^N \phi_\ell^2
		\label{eq:compat2}
		\end{equation}
		subject to the bound inequality constraints
		\[
		0\leq \phi_{\ell} \leq \widehat{(\phi_\ell)}_K,\quad \ell\in\{1,...,N\}
		\]
		and compatibility linear equality constraints
		\[
		\sum_{\ell=1}^N \phi_\ell\, (\nabla^h z_\ell)_K = \bm{0}.
		\]
		This optimization problem can be easily solved by standard duality 
		theory~\cite{optim}. We will do not detail the resulting algorithm.
		The construction may be a little more diffusive because of
		the double limitation process, but let us emphasize that areas with
		more than two materials are generally sparse, limited to singular
		topology elements like triple-points for example. 
		%
		%
		%
		%
		%
		\subsection{Numerical evaluation on the reference ``triple-point'' test case}
\begin{figure}[ht]
		\begin{center}
		\includegraphics[height=0.18\textheight]{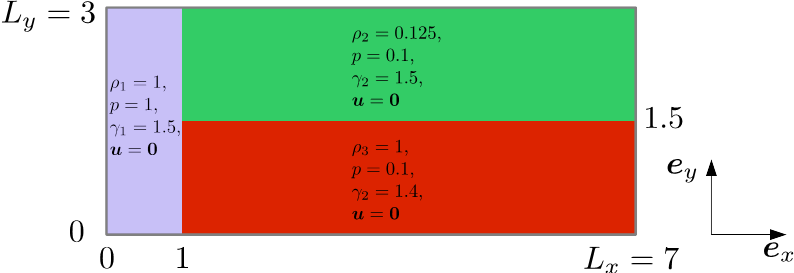}
		\caption{Geometry and initial configuration for the reference triple-point case.}
		\label{fig:triple}
		\end{center}
		\end{figure}

		%
		The resulting hydrodynamic solver is tested on the reference ``triple point''
		test case, found e.g. in Loub\`ere et al.~\cite{Loubere}. 
		This problem is a three-state two-material 2D Riemann problem in a rectangular vessel.
		The simulation domain is $\Omega=(0,7)\times(0,3)$ as described in figure~\ref{fig:triple}.
		The domain is splitted up into three regions $\Omega_i$, $i=1,2,3$ filled with two perfect gases leading to 
		a two-material problem. Perfect gas equations of state are used with 
		$\gamma_1=\gamma_3=1.5$ and~$\gamma_2=1.4$. Due to the density differences, two
		shocks in subdomains $\Omega_2$ and $\Omega_3$ propagate with different speeds. This
		create a shear along the initial contact discontinuity and the formation of a vorticity.
		Capturing the vorticity is of course the difficult part to compute. We use a rather
		fine mesh made of $2048\times 878$ points (about 1.8M cells).
		
\begin{figure}[ht]
		\begin{subfigure}{0.49\textwidth}
		\begin{center}
		\includegraphics[width=\linewidth]{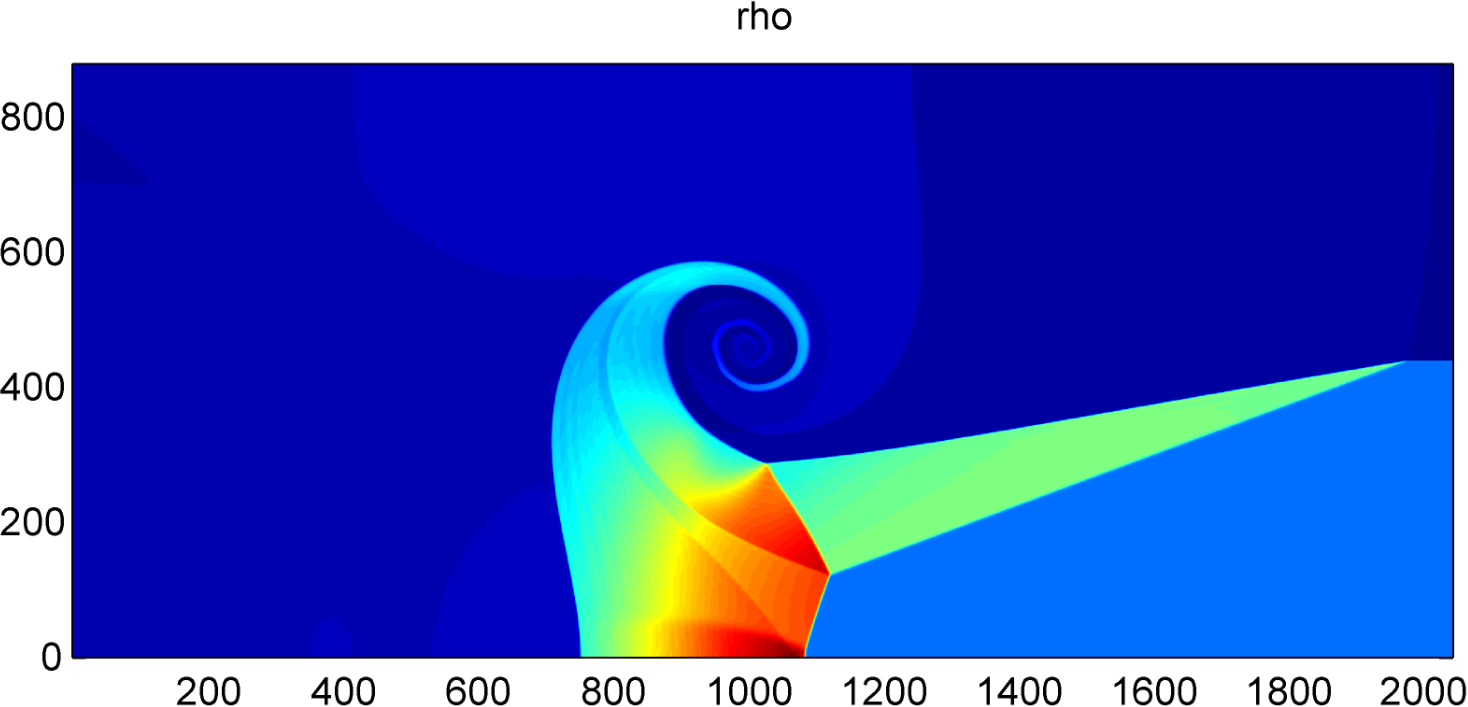}
		\end{center}
		\caption{Density field}
		\end{subfigure}
		\begin{subfigure}{0.49\textwidth}
		\begin{center}
		\includegraphics[width=\linewidth]{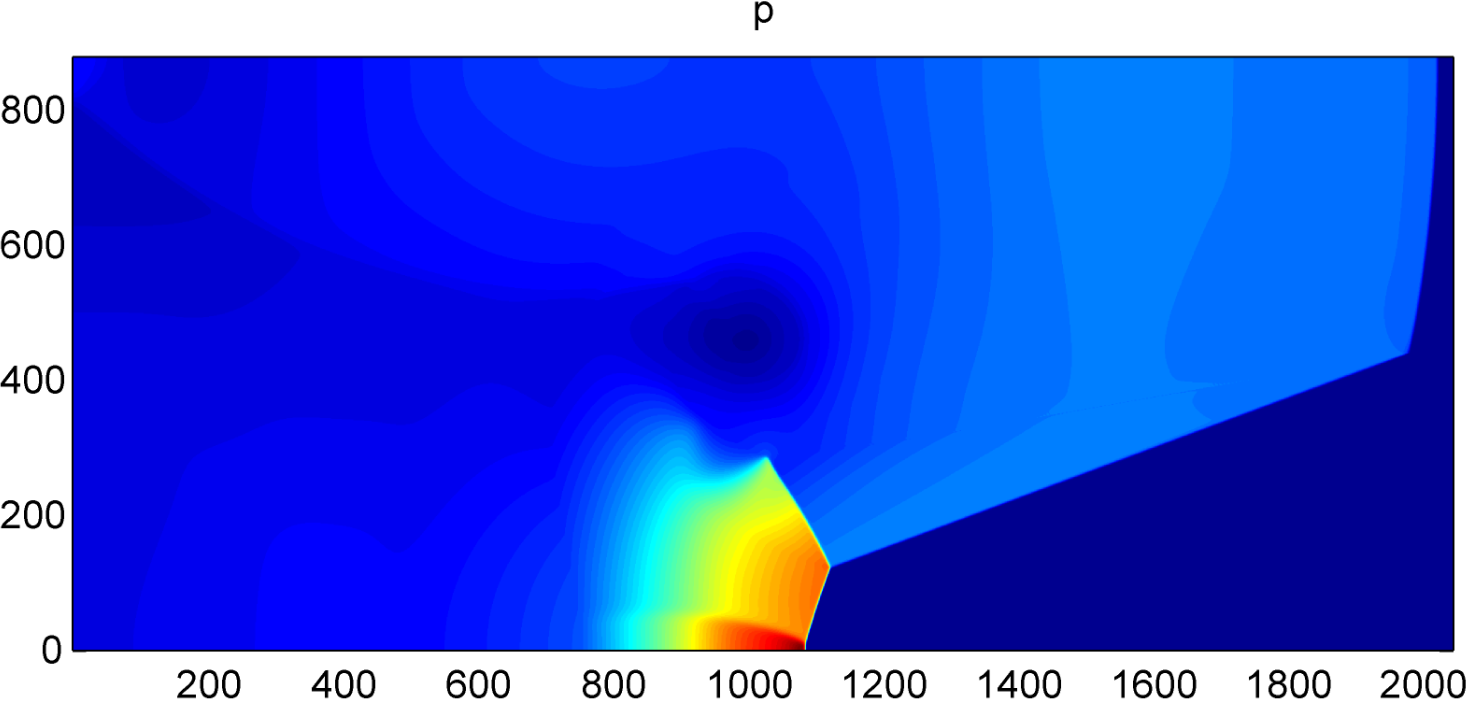}
		\end{center}
		\caption{Pressure field}
		\end{subfigure}
		\begin{subfigure}{0.49\textwidth}
		\begin{center}
		\includegraphics[width=\linewidth]{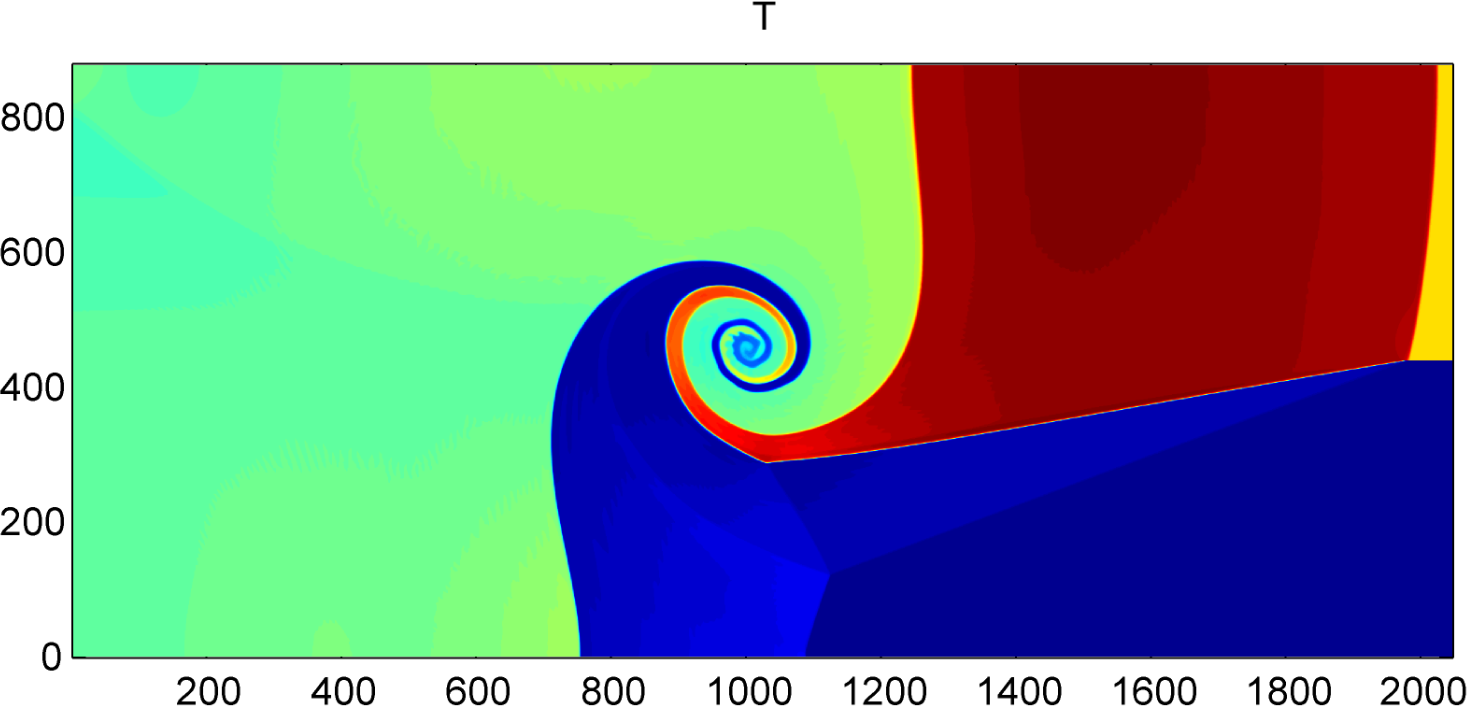}
		\end{center}
		\caption{Temperature field}
		\end{subfigure}
		\begin{subfigure}{0.49\textwidth}
		\begin{center}
		\includegraphics[width=\linewidth]{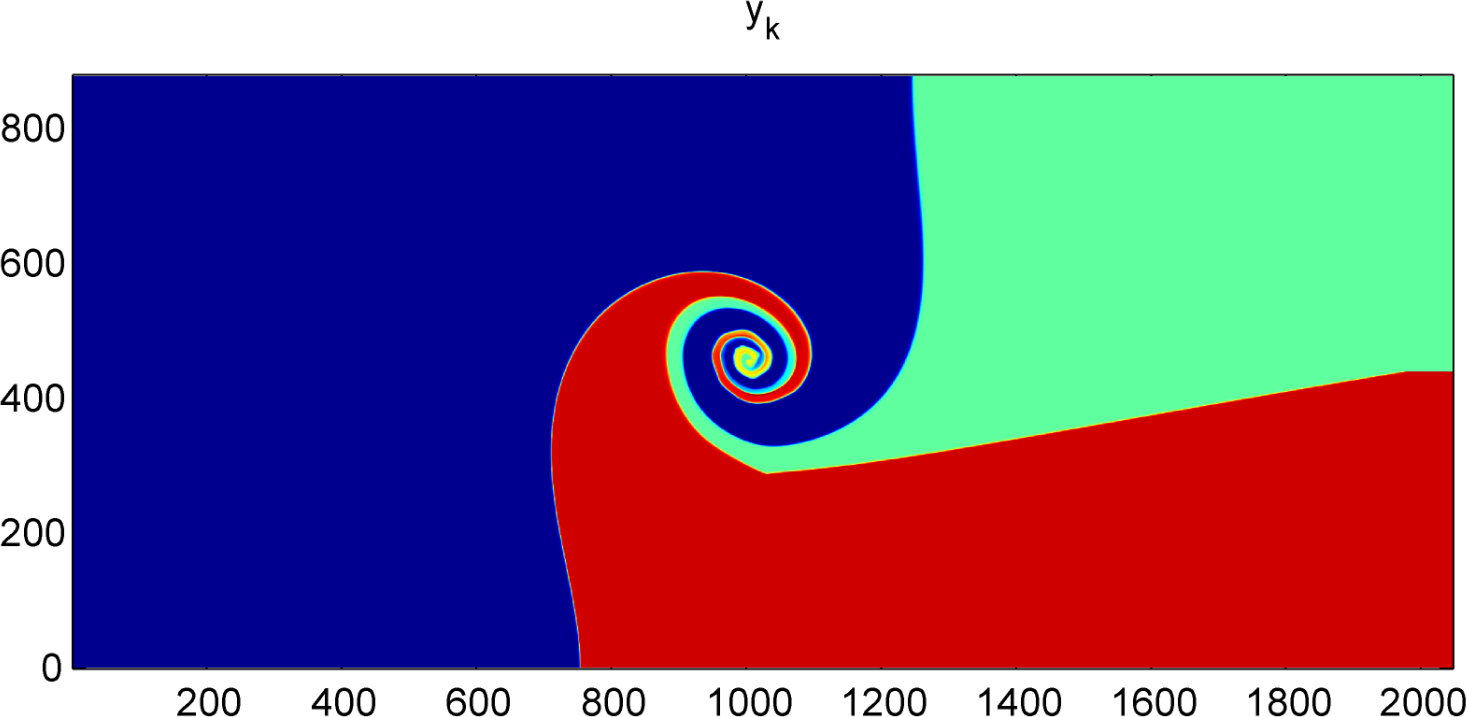}
		\end{center}
		\caption{Colored representation of material indicators}
		\end{subfigure}
		\caption{Results on the multimaterial ``triple point'' case  (perfect gases) 
		using a collocated Lagrange+remap solver 
		+ low-diffusive interface capturing advection scheme, mesh 2048x878.
		Final time is $T_f=3.3530$.}
		\label{fig:triple1}
		\end{figure}

		On figure~\ref{fig:triple1}, the numerical solution at final time $T_f=3.3530$. We also provide on figure~\ref{fig:2} a zoom of the vorticity zone to show the accuracy of the interface
		captures in this area.
		\begin{figure}[ht]
		\centering\includegraphics[width=0.6\textwidth]{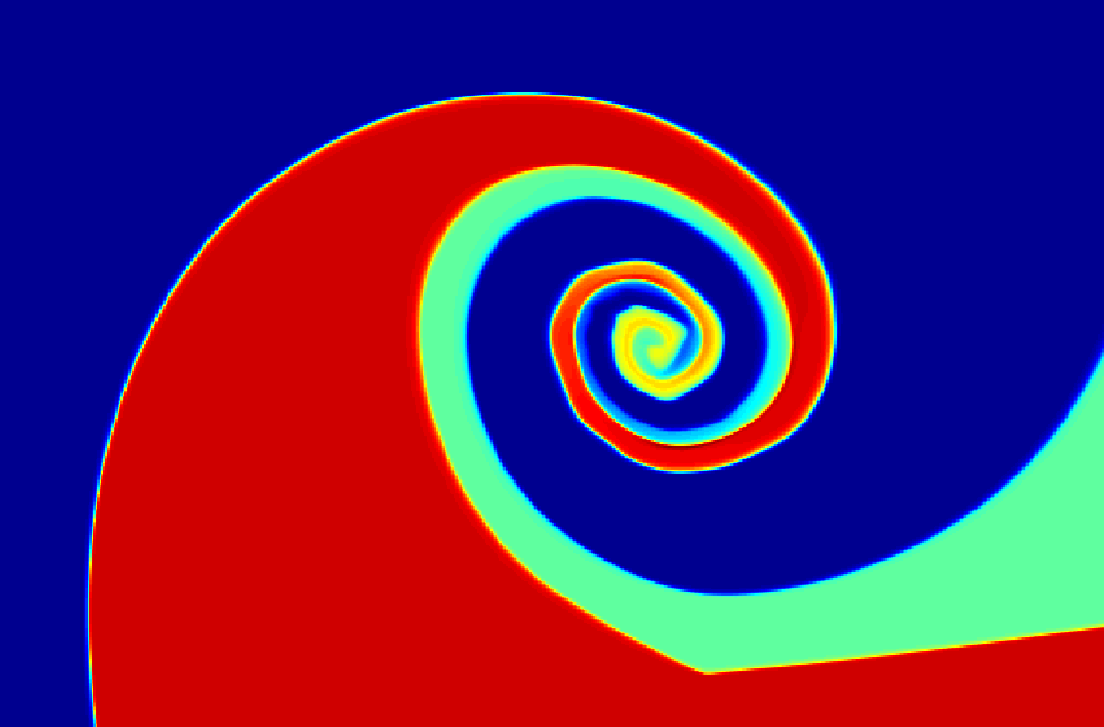}
		\caption{Triple point, zoom-in of the vortex region at final time, mesh 2048x878.}
		\label{fig:triple2}
	\end{figure}
\section{Concluding remarks and perspectives}
%
In this paper, topics on accuracy, stability and artefact-free interface capturing
methods have been discussed and analyzed. On this subject of high interest
for the whole Hydrodynamics community, of course one can already find important literature and contributions. We have tried to shed a new light and bring our understanding of
this tricky subject. Artefacts commonly encountered with interface capturing schemes (including strange attractor shapes like octagons on cartesian grids,
privileged directions artefacts, zigzag interface instabilities) are mainly due to
a misbalance between too much antidiffusion and expected accuracy,
but also because of important errors of gradient direction estimation. In particular
direction-by-direction compressive slope limiters (superbee, ultrabee, limited
downwind, ...) are too compressive in some directions and do not take into
account the direction normal to the interface which has to be evaluated accurately
as demonstrated in this paper. For this purpose, a multidimensional 
limiting process (MLP) strategy appears to be a good candidate that first evaluate
the gradient direction without any nonlinear limitation, and then limit the
gradient intensity in order to get local extremum diminishing (LED) properties.
Numerical evidence also shows that a first-order explicit Euler scheme 
creates linear instabilities that evolve toward spurious zigzag-shaped interface
modes when the numerical advective flux does not include the compensating
Lax-Wendroff term. Rather than including a Lax-Wendroff diffusive flux, we rather
use a Runge-Kutta RK2 time integrator that allows to kill zigzag instabilities.
The whole MLP+RK2 strategy provide a stable and accurate diffuse interface capturing 
approach with an acceptable discrete interface compactness.
Although it is not at the aim on paper, we wanted to have a first result of extension
to compressible multimaterial hydrodynamics to show both generalisation and
flexibility of the method and have a first idea of the competitiveness of
the method, especially compared to interface reconstruction algorithms. We believe
that this approach is promising in terms of accuracy and computing performance
(pure SIMD algorithms with natural parallelization on multicore/manycore architectures). We also intend to extend and evaluate the approach on unstructured
meshes.

\section*{Acknowledgements}
This work is part of the joint lab agreement (LRC MESO) between CMLA and CEA DAM DIF. The first author would like to thank J. Ph. Braeunig, D. Chauveheid, S. Diot and A. Llor for fruitful and
valuable discussions.
%

%
\end{document}